\documentclass[fleqn,usenatbib]{mnras}

\usepackage{newtxtext,newtxmath}

\usepackage[T1]{fontenc}

\DeclareRobustCommand{\VAN}[3]{#2}
\let\VANthebibliography\thebibliography
\def\thebibliography{\DeclareRobustCommand{\VAN}[3]{##3}\VANthebibliography}

\usepackage{graphicx}	
\usepackage{amsmath}	

\usepackage{xcolor}
\usepackage{mathtools} 
\usepackage{hyperref}
\usepackage{soul} 

\title[AGN Variability]{What Drives the Variability in AGN? Explaining the UV-Xray Disconnect Through Propagating Fluctuations}

\author[S. Hagen, C. Done, and R. Edelson]{
Scott Hagen,$^{1}$\thanks{E-mail: scott.hagen@durham.ac.uk}
Chris Done,$^{1}$
Rick Edelson$^{2}$
\\
$^{1}$Centre for Extragalactic Astronomy, Department of Physics, University of Durham, South Road, Durham DH1 3LE, UK\\
$^{2}$Eureka Scientific Inc., 2453 Delmer Street, Suite 100, Oakland, CA 94602, USA
}

\date{Accepted XXX. Received YYY; in original form ZZZ}

\pubyear{2023}

\newcommand{\rout}{r_{\mathrm{out}}}
\newcommand{\risco}{r_{\mathrm{isco}}}
\newcommand{\mdot}{\dot{m}}
\newcommand{\Mdot}{\dot{M}}
\newcommand{\MMdedd}{\dot{M}/\dot{M}_{\mathrm{Edd}}}
\newcommand{\Medd}{\dot{M}_{\mathrm{Edd}}}
\newcommand{\Ledd}{L_{\mathrm{Edd}}}

\newcommand{\fgen}{f_{\mathrm{gen}}}
\newcommand{\fprop}{f_{\mathrm{prop}}}
\newcommand{\Fvar}{F_{\mathrm{var}}}
\newcommand{\Ndec}{N_{\mathrm{dec}}}

\newcommand{\Msol}{M_{\odot}}
\newcommand{\LLedd}{L/L_{\mathrm{Edd}}}
\newcommand{\Ldiss}{L_{\mathrm{diss}}}
\newcommand{\Lseed}{L_{\mathrm{seed}}}
\newcommand{\kTh}{kT_{e, h}}
\newcommand{\kTw}{kT_{e, w}}
\newcommand{\fcov}{f_{\mathrm{cov}}}

\newcommand{\rwind}{r_\mathrm{wind}}
\newcommand{\hwind}{h_\mathrm{wind}}

\newcommand{\rev}{}

\begin{document}
\label{firstpage}
\pagerange{\pageref{firstpage}--\pageref{lastpage}}
\maketitle

\defcitealias{Ingram11}{ID11}
\defcitealias{Ingram12}{ID12}
\defcitealias{Hagen23a}{HD23a}

\begin{abstract}

\rev{
Intensive broadband reverberation mapping campaigns have shown that AGN variability is 
significantly more complex than expected from disc reverberation of the variable X-ray illumination. 
The UV/optical variability is highly correlated and lagged, with longer lags at longer wavelengths as predicted, but 
the observed timescales are longer than expected. Worse,  
the UV/optical lightcurves are not well correlated with the X-rays which should drive them. 
Instead, we consider an intrinsically variable accretion disc, where slow mass accretion rate fluctuations 
are generated in the optical-UV disc, propagating down to  
modulate intrinsically faster X-ray variability from the central regions. 
We match our model to Fairall 9, a well studied AGN with $L\sim 0.1L_{\mathrm{Edd}}$, where the spectrum is dominated by the UV/EUV.
Our model produces lightcurves where the X-rays and UV have very different fast variability, yet are well 
correlated on longer timescales, as observed. It predicts that the intrinsic variability has optical/UV {\it leading} the X-rays,
but including reverberation of the variable EUV from an inner wind produces a lagged bound-free continuum which matches 
the observed UV-optical lags.  We conclude that optical/UV AGN variability is
likely driven by intrinsic fluctuations within the disc, not X-ray reprocessing: the observed longer than expected  lags are produced by reverberation of the EUV illuminating a wind not by X-ray illumination of the disc: the increasing lag with increasing wavelength is produced by the increased contribution of the (constant lag) bound-free continuum to the spectrum, 
rather than indicating intrinsically larger reverberation distances for longer wavelengths. 
}

\end{abstract}




\section{Introduction}

Active galactic nuclei (AGN) are powered by accretion onto a supermassive black hole (SMBH), a process that is generally understood through the framework of a \citet{Shakura73} optically thick and geometrically thin disc. In this model the energy dissipated through the flow gives rise to a radial temperature profile, increasing at smaller radii. The full spectral energy distribution (SED) from this model is the sum of black-body components, peaking typically in the UV/EUV for bright AGN.

A key feature of the standard \citet{Shakura73} disc theory is that the time-scale for changing the mass-accretion rate through the disc is the viscous timescale, typically several thousand years for SMBH (see e.g \citealt{Noda18}). However, observations of optical/UV emission from AGN clearly show stochastic variability on time-scales of months to years (e.g. \citealt{MacLeod10}). This is incompatible with standard disc theory \citep{Lawrence18}.

However, AGN SEDs are more complex than a simple sum of black-body components. AGN spectra always include an X-ray tail extending to higher energies, showing that some energy is dissipated outside of the optically thick disc structure \citep{Elvis94, Lusso16}.
This X-ray emission is generally highly variable on short time-scales, showing it is formed in a compact corona (e.g \citealt{Lawrence87, Ponti12b}). 
This gives a potential solution of the variability problems as some fraction of the variable X-ray emission should illuminate the disc, giving a variable reprocessed component in the optical/UV 
\citep{Clavel92}. 
This can be directly tested by long, well sampled lightcurves in the X-ray, UV and optical as it predicts that the variations in the X-ray emission 'echo' through the disc, progressively modulating first the UV from the inner disc then the optical from the outer disc. The radial temperature profile of a standard disc predicts a relation for the time lag between variability of the X-ray source and that at any wavelength $\lambda$ emitted by the disc of  $\tau \propto \lambda^{4/3}$ \citep{Collier99, Cackett07}. The recent intensive multi-wavelength monitoring campaigns of AGN, 
were designed to use this relation to 
map the accretion disc size scale \citep{Edelson15, Edelson17, Edelson19, McHardy14, McHardy18, Fausnaugh16, Chelouche19, Cackett18, Hernandez20, Kara21, Vincentelli21}. 

However, what these campaigns have instead showed is that AGN variability is significantly more complex than predicted by these models. Generally the measured lags give a size scale several times larger than expected for a disc, and often the correlation between the X-ray and UV/optical is poor. The long lags can be produced if there are additional structures such as e.g, a wind on the inner edge of the BLR contributing to the re-processed signal (e.g \citealt{Dehghanian19b, Chelouche19, Kara21}), \rev{or if the X-ray source is located at larger scale heights above the black-hole than expected by gravity \citep{Kammoun21a, Kammoun21b}}. The poor correlation between the X-ray and UV/optical, however, cannot easily be explained in the standard reverberation picture. Attempts at directly modelling the UV/optical light-curves through disc reverberation all predict light-curves that are highly correlated with the X-ray \citep{Gardner17, Mahmoud20, Mahmoud23, Hagen23a}. Reverberation smooths on timescales similar to the lags, and the optical/UV lags are of the order a few days, yet these lightcurves have typical variability timescale of 20-40 days, while the X-rays have typical timescale of $<1$ day. This amount of smoothing cannot be produced by the same reverberation material which gives rise to the lags, strongly indicating that reverberation is not the sole driver of variability in AGN.

Intrinsic variability of the accretion flow is even more directly required in the 'changing look' AGN. These 
show a transition in Balmer line profiles, from type 1 (broad plus narrow lines) to type 1.8-2 (where the broad component of the line is strongly suppressed). This change correlates with a drop in the observed optical/UV  continuum flux. Most of these events are not likely due to obscuration, as posited by unification models, as there is no associated reddening signature \citep{LaMassa15, McElroy16, Ruan16, Runnoe16}. Even more convincingly, the infrared emission (reprocessed UV from the torus) also follows the optical/UV, showing clearly that this is an intrinsic change in the AGN accretion flow \citep{Sheng17, Stern18, Wang18, Ross18}, likely marking the transition between some sort of efficient optically thick disc accretion, to being dominated by a radiatively inefficient optically thin hot flow \citep{Noda18,Ruan19}.

In this paper we will consider the scenario where the accretion disc itself is intrinsically variable on observable time-scales. This obviously departs from standard \citet{Shakura73} disc theory. However, 
the disc emission itself also clearly departs from standard \citet{Shakura73} theory in its disc spectral shape \rev{(e.g \citealt{Antonucci89, Lawrence18})}. There is a ubiquitous downturn in the far UV which appears to extrapolate 
across the unobservable EUV data gap to meet an upturn below 1~keV relative to the X-ray coronal tail \citep{Laor97, Porquet04, Gierlinski04} (historically referred to as 'the Big Blue Bump' and soft X-ray excess). These two features can be modelled together with a single warm, optically thick thermal Compton emission component \citep{Mehdipour11, Mehdipour15, Done12, Kubota18, Petrucci18}. 
Such models can provide successful fits to the optical/UV/X-ray SEDs of individual AGN (e.g \citealt{Matzeu16, Done16, Czerny16, Hagino16, Hagino17, Porquet18, Porquet19}) and also larger samples of AGN (e.g \citealt{Jin12a, Jin12b, Mitchell23, Temple23}). 

The warm Comptonisation component generally dominates the bolometric luminosity, indicating its origin is in the energy generating structure. It also requires a 
large optical depth. Both these strongly suggest that this is emission from the disc itself, but obviously it is not thermalising to a blackbody as in the \citet{Shakura73} models. Instead, one way to produce the spectrum is to change the vertical structure of the disc such that the accretion power is 
dissipated higher in the photosphere 
\citep{Rozanska15, Petrucci18, Jiang20}, rather 
than concentrated on the equatorial plane as in the  
\citet{Shakura73} models. However, this is not yet well 
understood, so there are no theoretical models to predict 
the intrinsic variability of this structure, though this is the subject of recent numerical studies \citep{Secunda23b}.
Instead, we will use the observations as a guide, and 
model the intrinsic variability phenomenologically. 
The results can then be used to build more physical models of the disc structure, as well as to give more realistic driving lightcurves to reverberation map the accretion flow structures.  

We take models of the disc and corona variability seen in the stellar mass black hole binaries as our starting point. 
Unlike the AGN, these can show spectra which are dominated by what looks very like a standard \citet{Shakura73} disc (high/soft state see e.g. \citealt{Done07}, X-ray tail is very weak). Again unlike AGN, the variability of this component is well matched to the viscous timescale of the outer disc of weeks/months \citep{Dubus01, Lasota07, Lasota08}, with no short timescale disc variability. However, they also show spectra (bright low/hard and intermediate states) with more X-ray tail
where the disc emission is much weaker and distorted from the \citet{Shakura73} models. This disc emission
varies on timescales of seconds \citep{Uttley11, DeMarco17,Kawamura22}, much faster than 
expected from a viscous timescale even from the inner disc. This may instead indicate a turbulent region on the inner edge of a truncated disc, where the flow transitions to a hot corona 
\citep{Kawamura22, Kawamura23, Marcel22, Lucchini23}.

\begin{figure}
    \centering
    \includegraphics[width=\columnwidth]{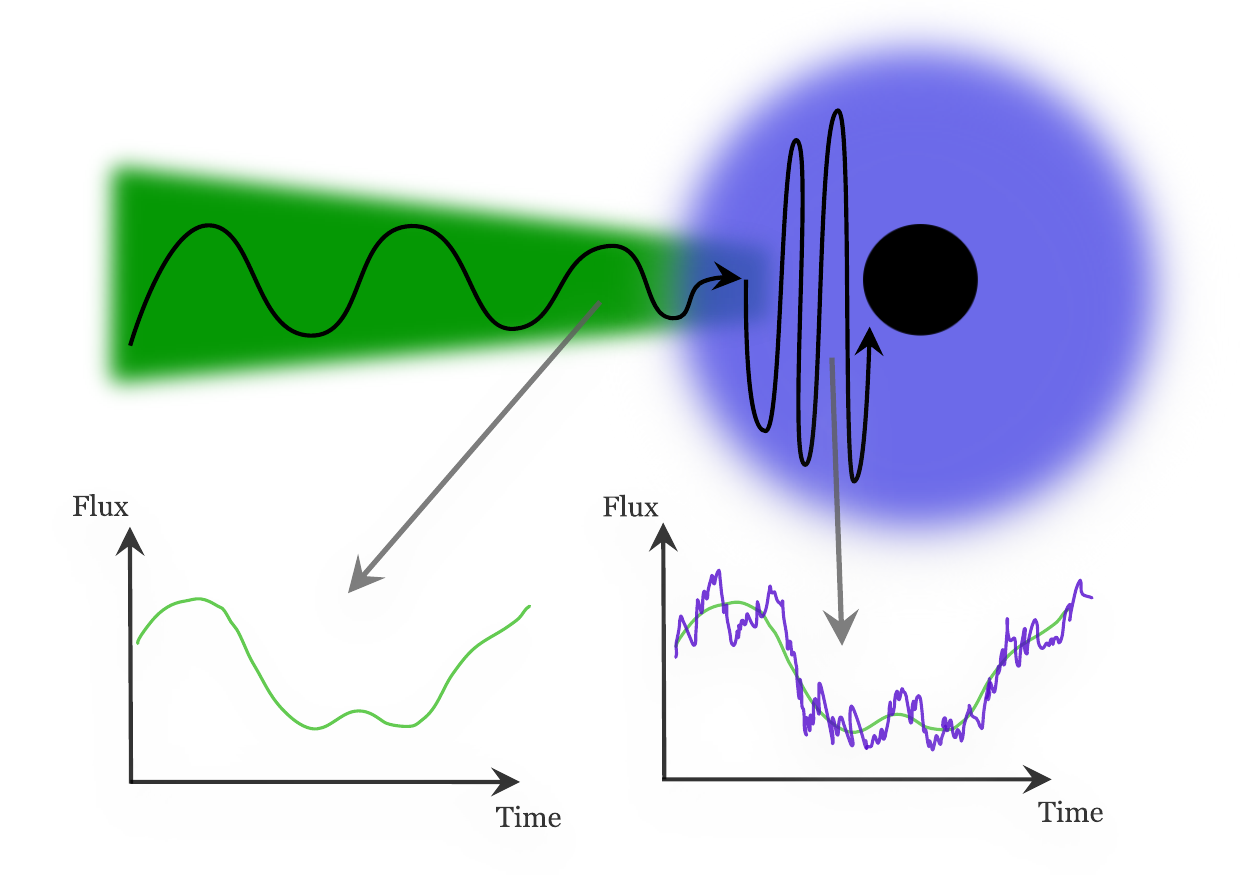}
    \caption{A depiction of what we envisage with the propagating fluctuations. The disc like structure (green) generates slow variability, giving light-curves that vary on long time-scales (bottom left). These variations propagate into the hot Corona (blue), which generates fast variability. The slow variations from the disc modulate the fast from the corona, giving light-curves containing both long and short term trends (bottom right). We stress that this is a sketch, and as such the components and light-curves are not to scale.}
    \label{fig:prop_sketch}
\end{figure}

The first models of the variability in stellar mass systems only focused on the hard X-ray tail, which shows the fastest variability with substantial power down to $\sim 0.1$~s. The broad band power spectrum, and lags between different energy bands in the X-ray power law emission can be reproduced in models based on 
a propagating fluctuations framework.
Larger radii typically produce longer timescale fluctuations, and these drift inwards, modulating the shorter timescale fluctuations stirred up at smaller radii \citep{Lyubarskii97, Arevalo06, Ingram11, Ingram12}. 
These models were then
extended to include the turbulent disc region. The key feature is that there is a discontinuity in timescale of intrinsic fluctuations at the radius at which the geometrically thin, cool disc transitions into the geometrically thick hot flow \citep{Rapisarda16, Mahmoud18b, Kawamura22, Kawamura23}. Since fluctuations propagate inwards, the hard X-ray lightcurve contains the propagated slow fluctuations from the disc, as well the faster fluctuations generated in the corona itself, while the intrinsic disc lightcurve contains only the slow fluctuations generated within the disc, as shown schematically in Figure \ref{fig:prop_sketch}.
This clearly holds out the possibility to explain the disconnect in variability timescale seen in AGN, where the disc lightcurve is intrinsically much smoother 
than the X-ray lightcurve. This also retains a correlation between disc and X-ray lightcurves on long timescales due to the propagation of the disc fluctuations, but gives an overall poor correlation on faster timescales as these are produced only in the X-ray hot flow so are not intrinsically part of the disc variability. 

We will scale up these models from black hole binaries to AGN, using Fairall 9 as a typical example of a moderate luminosity Seyfert 1 galaxy. This AGN was the subject of a long timescale intensive broadband monitoring campaign, giving excellent spectral and variability data \citep{Hernandez20}, that show outgoing variations on short time-scales as well as tentative evidence of propagation on long time-scales where the lag time is 10s of days \citep{Hernandez20, Neustadt22, Yao23}.

We first build a broadband spectral model which is tailored for variability studies (Section 2) and then parameterise propagating fluctuations in mass accretion rate through this structure (Section 3). This full spectral-timing model predicts the intrinsic lightcurves in any energy band, so gives input for reverberation (Section 4). 
We allow the X-ray emission to illuminate the disc to produce a disc reverberation signal, and use the bolometric (especially extreme UV) flux to
reverberate from a larger scale wind on the inner edge of the BLR. 
Finally, in Section 5 we generate a set of model light-curves and explore how the different components within the accretion flow affect the observables (i.e lags and cross-correlation functions \rev{(hereafter CCF)}).

\section{Modelling the SED}

Throughout this paper we use $R$ for radius in physical units (measured from the black hole) and $r$ for dimensionless gravitational radii, where $R = rR_{G}$ with $R_{G} = GM/c^{2}$. Additionally, we will use $\mdot$ for dimensionless mass accretion rate, scaled by the Eddington rate such that $\mdot = \MMdedd$, where $\Mdot$ is the physical mass-accretion rate in $g/s$, and $\Medd$ is the Eddington mass-accretion rate, related to the Eddington luminosity by $\Ledd = \eta(a)\Medd c^{2}$, where $\eta(a)$ is a spin-dependent accretion efficiency.

We base our spectral-timing model on the underlying radially stratified SED 
{\sc agnsed} from \citet{Kubota18}.
This assumes the radial 
emissivity profile for a thin disc, $\epsilon(R) \propto R^{-3} f(R)$ where $f(R)$ describes the stress free inner boundary condition for the Kerr metric \citep{Novikov73, Page74}, but allows the emission from the flow to 
transition from blackbody from a standard outer disc ($r > r_{w}$), to warm Comptonisation ($r_{w} \geq r > r_{h}$), and then to hot Comptonisation ($r_{h} \geq r > \risco$). Within our code, the standard outer disc and warm Comptonisation region are implemented as in \citet{Kubota18}, and we refer the readers to this paper for details regarding these two regions. For the hot Comptonisation region, however, we make some small modifications to better follow the spectral variability. 

\begin{figure}
    \centering
    \includegraphics[width=\columnwidth]{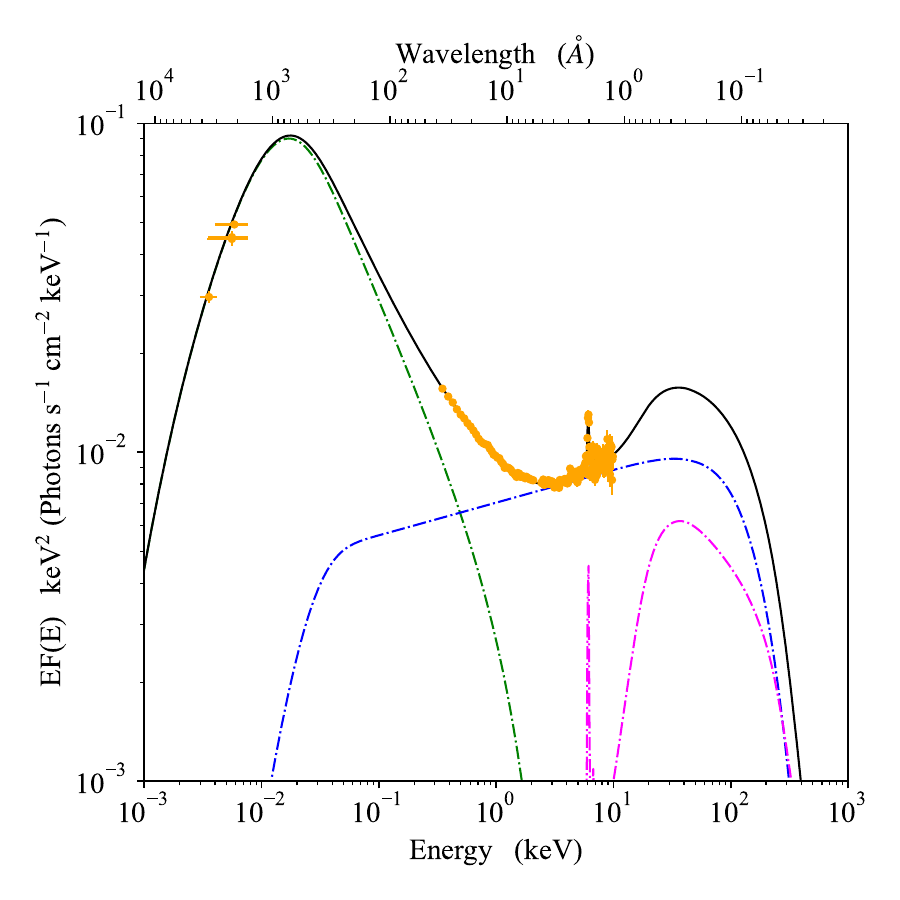}
    \caption{The mean SED of Fairall 9 during the 1st year of the intensive monitoring campaign, which will be used throughout this paper to initiate the variable model. The solid black line shows the total {\sc $\tau$agnsed} model, with the components shown separately as dashed lines. The hot Compton flow is in blue with its neutral reflection in magenta.  The whole of the optically thick disc emission is modelled as     
    warm Compton (green dashed line) rather than including any standard \citet{Shakura73} disc region. This is 
    important later when we consider the variability time-scales of the system.}
    \label{fig:f9SED}
\end{figure}

The hot Compton emission in {\sc agnsed}  is parameterised by the coronal radius $r_{h}$, photon index $\Gamma_{h}$, and electron temperature $kT_{e, h}$. However, the photon index and electron temperature are fundamentally set by the balance between Compton heating and cooling per electron within the hot flow (see e.g. \citealt{done10}). Compton heating depends on the power, $\Ldiss$, dissipated within $r_h$ while cooling depends on the seed photon luminosity from the disc which is incident on the hot flow, $\Lseed$ (see \citealt{Hagen23b} for an update on how $\Lseed$ is calculated). Following 
\citet{Beloborodov99} (see also the {\sc qsosed} model in \citealt{Kubota18}) this gives:

\begin{equation}
    \Gamma_h = \frac{7}{3} \left( \frac{\Ldiss}{\Lseed} \right)^{-0.1}
    \label{eqn:gamma}
\end{equation}

We use this formalism to better capture the time dependent behaviour of the X-ray spectrum as changes in the seed photons from the disc $\Lseed$ travel at the speed of light, whereas mass accretion rate fluctuations which modulate $\Ldiss$ propagate more slowly (see e.g. \citealt{Veledina16, Veledina18}). This means that the power law spectrum pivots rather than changing only in normalisation \citep{Mastroserio18,Uttley23}. 

Continuing from \citet{Beloborodov99}, we can then calculate the coronal electron temperature, $\kTh$, from $\Gamma_h$ and $\tau_{h}$:

\begin{equation}
    \frac{\kTh}{ m_{e} c^{2}} = \frac{4 y}{\tau_{h}(\tau_{h} + 1)}\ \ {\rm where}\ y = \left( \frac{4 \Gamma_{h}}{9} \right)^{-9/2}
    \label{eqn:kth}
\end{equation}
where $y$ is the Compton y-parameter. 

This gives an SED model where $\Gamma_{h}$ and $\kTh$ are calculated self-consistently, and can therefore become time-dependent when we calculate the variable SED. We call this model {\sc $\tau$agnsed}.

\subsection{Fairall 9 fit to {\sc $\tau$agnsed}.}

{\renewcommand{\arraystretch}{1.6} 
\begin{table}
    \centering
    \begin{tabular}{c|c|c}
    \hline
    Component & Parameter (Unit) & Value \\
    \hline
    {\sc phabs} & $N_{H}$ ($10^{20}$\,cm$^{-2}$) & 3.5 \\
    \hline
    {\sc $\tau$agnsed}  & $M$ ($\Msol$) & $2\times 10^{8}$\\
                        & Dist (Mpc) & 200 \\
                        & $\log \mdot$ ($\MMdedd$) & $-1.217^{+0.0196}_{-0.024}$ \\
                        & $a_{\star}$ & $0.722^{+0.062}_{-0.11}$ \\
                        & $\cos(i)$ & 0.9 \\
                        & $\tau_{h}$ & $0.99^{+2.4}_{-0.99}$ \\
                        & $\kTw$ (keV) & $0.394^{+0.037}_{-0.036}$ \\
                        & $\Gamma_{w}$ & $2.821^{+0.028}_{-0.029}$ \\
                        & $r_{h}$ & $9.16^{+0.90}_{-0.81}$ \\
                        & $r_{w}$ & $=\rout$ \\
                        & $\log \rout$ & \rev{$=r_{sg}=530$} \\
                        & $h_{\mathrm{max}}$ & $10$ \\
                        & Redshift & $0.045$ \\
    \hline
    {\sc rdblur}    & Index & -3 \\
                    & $r_{\mathrm{in}}$ & $386^{+517}_{-163}$ \\
                    & $r_{\mathrm{out}}$ & $10^{6}$ \\
                    & Inc (deg) & 25.8 \\
    \hline
    {\sc pexmon}    & $\Gamma$ & Calculated from {\sc $\tau$agnsed} \\
                    & $E_{c}$ (keV) & $10^{4}$ \\
                    & Redshift & $0.045$ \\
                    & Inc (deg) & 25.8 \\
                    & Norm ($10^{-3}$) & $4.52^{+0.70}_{-0.60}$ \\
    \hline
    \hline
    $\chi^{2}$/d.o.f & 229.50/166 = 1.38 \\
    \hline
    \end{tabular}
    \caption{Fit parameters for the mean SED of Fairall 9. Values with no error were frozen during the fitting process. We note that the inner radius in {\sc rdblur} was kept free to fit the Fe-K$\alpha$ line profile which may have substantial contribution from material further out in the accretion flow (wind/BLR and torus). 
    The lower limit on $\tau_{h}$ is formally $0$ as this is degenerate with $kT_{e,h}$ and our data do not have sufficient spectral coverage to constrain the high energy rollover. \rev{Finally, the outer radius is fixed at the self-gravity radius, calculated from \citet{Laor89}}. This forms our mean SED model throughout the paper when we calculate the variability.}
    \label{tab:f9SED}
\end{table}
}

We refit the Fairall 9 SED from \citet{Hagen23a} (hereafter \citetalias{Hagen23a}) with this new model, so as to get the time averaged values of each parameter, using {\sc xspec} v.12.13.0c \citep{Arnaud96}. \rev{We assume a standard Cosmology of $H_{0}=69.9$, $\Omega_{m}=0.29$, and $\Omega_{\nu} = 0.71$ for a flat Universe.}  As in \citetalias{Hagen23a}, we also include a neutral reflection component to account for the Fe-K$\alpha$ line and Compton hump, modelled with {\sc pexmon} \citep{Nandra07, Magdziarz95} convolved with {\sc rdblur} \citep{Fabian89} to account for any smearing within the reflection spectrum. We tie the value of the hard X-ray spectral index in {\sc pexmon} to that in {\sc $\tau$agnsed}. 
The optical/UV fluxes are dereddened, and host galaxy subtracted (see \citealt{Hernandez20} for details), so we include
galactic absorption only on the X-ray data, using {\sc phabs} with fixed 
$N_{H} = 0.035 \times 10^{22}$\,cm$^{2}$.
The final {\sc xspec} model is then {\sc phabs*($\tau$agnsed + rdblur*pexmon)}. The resulting intrinsic
model SED is shown compared to the 
data (unabsorbed, and deconvolved from the instrument response) in Fig. \ref{fig:f9SED}, with the corresponding fit parameters given in Table. \ref{tab:f9SED}.

In common with many intensively studied AGN (e.g. \citealt{Mehdipour11,Mehdipour16,Petrucci18}, there is no need for an outer standard disc component. The flow is well described using only the warm Comptonised disc from $r \sim 500-10$ and a hot inner flow from $10-\risco$. \rev{The properties of the X-ray corona calculated by the model are $\Gamma_{h}=1.9$ and $\kTh=83.1$\,keV.} We will use these parameters from the time averaged SED to set the size scale and physical conditions for our model of 
variability from propagating fluctuations through this structure. 

\section{Propagating fluctuations in {\sc $\tau$agnsed}}

\subsection{Propagating Fluctuations}

\begin{figure*}
    \centering
    \includegraphics[width=\textwidth]{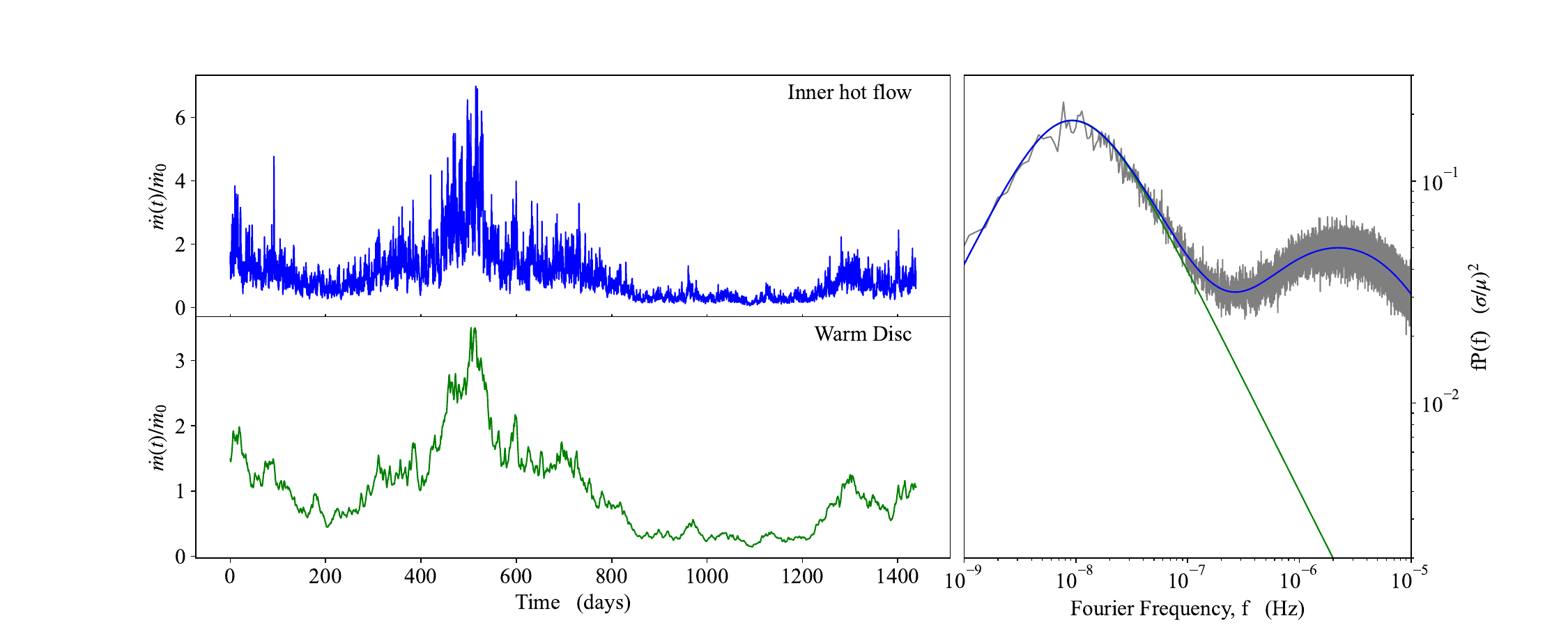}
    \caption{Example mass-accretion rate time-series and power-spectrum for a two component flow, consisting of a slowly varying warm disc region (green) and an intrinsically faster hot corona (blue). \\
    {\it Left}: $\sim 1500$ day snapshot from a time-series realisation calculated for $N=2^{18}$ time-steps and sampling rate $dt = 0.1$ days. It is clear that the time-series from within the hot corona (blue) has both stronger and faster variability than the disc. \\
    {\it Right}: The power-spectrum for the inner part of the warm disc (green) and the inner part of the hot flow (blue). The coloured lines show the analytic solution, following \citet{Ingram13}, while the grey line shows the average power-spectrum from 100 time-series realisations, again using $N=2^{18}$ and $dt=0.1$ days. As expected, these follow the analytic treatment. It can be seen that the hot coronal variability closely follows that of the warm disc at low frequencies, due to the disc variations propagating into the corona, and hence modulating its variability.}
    \label{fig:example_mdot_powSpec_and_tseries}
\end{figure*}

The propagating fluctuations model considers local variations in the mass accretion propagating down through the accretion flow
\citep{Lyubarskii97}. 
In this work we will consider these propagating mass accretion rate fluctuations as the driver of intrinsic variability in AGN. The formalism used is based on \citet{Ingram11, Ingram12, Ingram13} as used in \citet{Kawamura23}, and we refer the reader to these papers for a detailed description of the model. For completeness, however, we give a brief overview of the key aspects here.

Following \citet{Ingram13} we split the accretion flow into $N$ annuli centred at $r_{n}$ with equal logarithmic spacing such that $d\log(r_{n}) = dr_{n}/r_{n}=\mathrm{constant}$. The local mass accretion rate variability is then assumed to have a well defined power-spectrum, given by a zero-centred Lorentzian peaking at the frequency at which fluctuations are locally generated, $\fgen$.

\begin{equation}
    |A(r_{n}, f)| = \frac{\sigma^2}{\pi T} \frac{\fgen(r_{n})}{(\fgen(r_{n}))^{2} + f^{2}}
\end{equation}
where $f$ is the Fourier frequency, and $\sigma$ and $T$ are the variance and duration of the time-series $a(r_{n}, t)$. The fractional root mean square variability is 
$\sigma/\mu = \Fvar/\sqrt{\Ndec}$, where $\mu$ is the mean, fixed at unity, $\Ndec$ is the number of radial bins per decade, and so $\Fvar$ is the fractional variability produced per decade in radius in the flow. Using $|A(r_{n}, t)|$ we can now create realisations of the time-series $a(r_{n}, t)$ using the method described in \citet{Timmer95}. Since the fluctuations generated within each annulus propagate down through the flow, the total time-series within each annulus will be modulated by those from the previous annuli. As this is a multiplicative process \citep{Ingram13}, we can write the mass-accretion rate time series at each annulus as:

\begin{equation}
    \mdot(r_{n}, t) = \mdot_{0} \prod_{k=1}^{n} a(r_{k}, t-\Delta t_{kn})
\end{equation}
where $\mdot_{0}$ is the mean mass-accretion rate, set to 1 throughout, and $\Delta t_{kn}$ is the propagation time between the annuli at $r_{k}$ and $r_{n}$.

In a standard \citet{Shakura73} disc, $\fgen$
is given by the viscous timescale in the flow.
However, this is clearly many orders of magnitude 
too slow to describe the data (see Introduction).  Instead, following \citet{Kawamura23}, we 
parameterise $\fgen$ as a broken power law, with 
a discontinuity at $r_h$ to allow for the much faster timescales expected in the hot flow compared to the warm disc.

\begin{equation}
    \fgen(r) = \begin{cases}
  B_{g,h} r^{-m_{g,h}} f_{K}(r)  &  r<r_h \\
  B_{g,w} r^{-m_{g,w}} f_{K}(r)  &  r>r_h 
\end{cases}
    \label{eqn:fgen}
\end{equation}
where $f_{K} = (1/2\pi)r^{-3/2}$ is the Keplerian frequency at $r$ in units of $c/R_{G}$.

We tailor our fiducial model for variability to Fairall 9.
From the SED fits above, we see that the warm disc structure extends from $r\sim 200-10$, i.e. spans more than an order of magnitude. However, the Fairall 9 optical and UV lightcurves all have similar variability timescales \citep{Hernandez20}, so we fix $m_{g,w}=-3/2$ so as to give constant fluctuation timescales across the entire warm disc. Conversely, the hot Compton region barely spans a factor 2 in radius, but there is 
some evidence for radial stratification of timescale across the hot flow seen in AGN power spectra, with more high frequency variability at higher X-ray energies \citep{Ponti12b, Ashton22,Tortosa23}. Hence we choose $m_{g,h}=1$ so that there is a factor $\sim 5$ increase in $\fgen$ with decreasing radius across the hot flow. There are $\sim 3$ years of well sampled optical/UV lightcurve for Fairall 9 showing variability, so we choose a generator timescale\rev{, $1/\fgen$,} of $\sim 3$~years for the warm disc (Edelson et al (sub)). 
The X-ray shows substantially more fast variability, extending with large fluctuation power up to around $0.1$~day (\citealt{Markowitz03, Markowitz04, Lohfink14}; \citetalias{Hagen23a}), so we choose a generator timescale of 0.1~day for the inner edge of the corona at $r\sim 4$. 

The propagation timescale between each annulus is characterised in terms of a propagation frequency $\fprop$
such that 
$\Delta t_{kn} = (dr_{n}/r_{n}) (1/\fprop(r_{n}))$ (e.g \citealt{Ingram13}). Early models had this set to the same as the generator timescale, but there is clear evidence in the BHBs that propagation of fluctuations happens on faster timescales than this (see e.g. \citealt{Kawamura23}). There are now some tentative detections of propagation in AGN disc lightcurves, including in Fairall 9, where the lag time is 10s of days \citep{Hernandez20,Vincentelli22, Neustadt22, Yao23}. Hence 
we set the propagation timescale to be $\fprop(r)=100\fgen(r)$ i.e. we keep the same power law indices, but increase the normalization, so that $B_{p,w}=100B_{g,w}$ and $B_{p,h}=10B_{g,h}$ (see \citealt{Kawamura23}). 
This gives a $10$ day propagation timescale from $r=100$ to $50$, which are typical radii at which the disc emission peaks in the V and UVW2 band, respectively, and a 20-40 day lag for fluctuations in UVW2 to propagate down into the hard X-ray corona. 

{\renewcommand{\arraystretch}{1.6} 
\begin{table}
    \centering
    \begin{tabular}{c|c}
        \hline
        Parameter & Value \\
        \hline
        \multicolumn{2}{|c|}{- - - - - - - Generator Parameters - - - - - - -} \\
         $r_{\mathrm{var, max}}$ & 200 \\
         $F_{\mathrm{var}, w}$ & 0.6 \\
         $F_{\mathrm{var}, h}$ & 0.5 \\
         $B_{g, w}$ & $5 \times 10^{-5}$ \\
         $B_{g, h}$ & 1 \\
         $m_{g, w}$ & -3/2 \\
         $m_{g, h}$ & 1 \\

        \multicolumn{2}{|c|}{- - - - - - - Propagation Parameters - - - - - - -} \\
        $B_{p, w}$ & $5 \times 10^{-3}$ \\
        $B_{p, h}$ & 100 \\
        $m_{p, w}$ & -3/2 \\
        $m_{p, h}$ & 1 \\

        \hline
    \end{tabular}
    \caption{Parameter values used for the generator and propagator frequencies (see Eqn. \ref{eqn:fgen}) for all simulation runs in this paper. Parameters denoted with the subscripts $w$ and $h$ correspond to the warm and hot corona respectively, while the subscripts $g$ and $p$ correspond to generative and propagation time-scales respectively.}
    \label{tab:varpars}
\end{table}
}

Fig. \ref{fig:example_mdot_powSpec_and_tseries}a 
shows example time-series for the mass-accretion rate fluctuations
using these parameters (see 
table \ref{tab:varpars}). We discard the first 200 days to allow propagation throughout the accretion flow. 
The lower panel shows the variability propagating through the inner edge of the warm disc, while the upper panel shows the total variability in mass accretion rate through the inner edge of the hot flow.  It is clear that the hot flow lightcurve has more fast variability than the warm disc, but that the two are correlated on long timescales. 

Fig. \ref{fig:example_mdot_powSpec_and_tseries}b shows the power spectrum of the mass accretion rate fluctuations propagating through the inner edge of the hot flow (grey). As the model is stochastic these data are the power-spectrum averaged over 100 realisations, but we also show the analytic solution of the propagating fluctuation model \citep{Ingram13}. The blue line shows 
the result for propagation through the entire flow, while the green line shows the result for propagation only through the warm disc. 
The warm disc has 
strong, but slow, variability 
while the hot corona has this variability propagated into its mass accretion rate fluctuations, plus much faster variability generated in the hot flow itself. Plainly this captures some aspects of the observed disconnect in timescales between the UV and X-ray behaviour seen in Fairall 9, so we use these parameters in our exploration of how this affects the spectrum as a function of time in the next section.

\subsection{Converting $\mdot(t, r)$ to Light-Curves via the SED model}

\begin{figure*}
    \includegraphics[width=\textwidth]{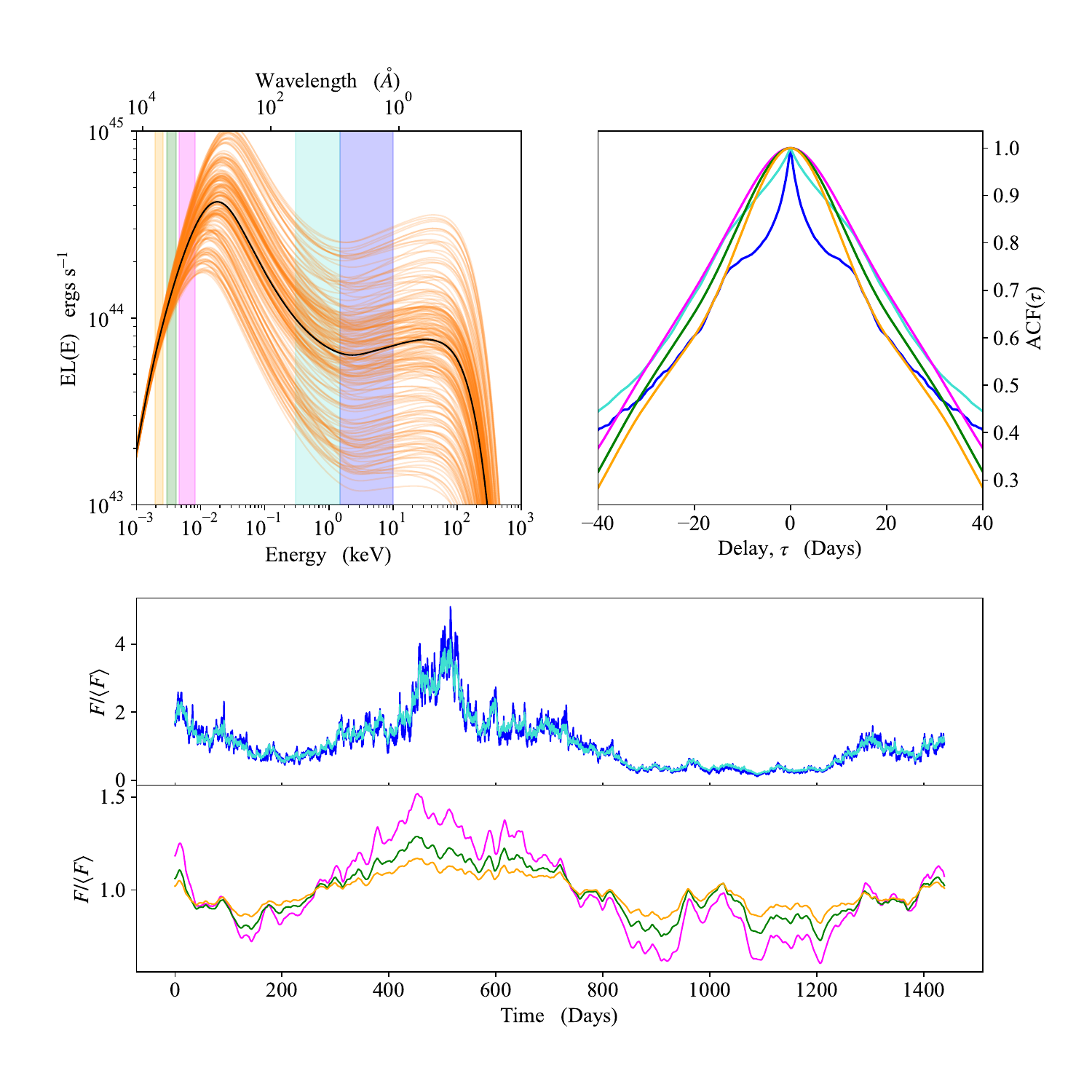}
    \caption{Example model output for a run considering the intrinsic variability only with no suppression of the disc variability seen by the hot corona, using the $\mdot$ realisation from Fig. \ref{fig:example_mdot_powSpec_and_tseries}.\\
    {\it Top Left}: The output SEDs. The solid black line shows the \rev{input SED calculated from the SED fit in section 2.1}, while the orange lines shows a sample of 200 SEDs randomly selected from the variable output. The shaded regions show the extraction regions used to generate light curves in HX (blue), \rev{SX (cyan)}, UVW2 (magenta), U (green) and (V) orange. For UVW2, U, and V the Swift-UVOT effective area curves were used to calculate the light-curves. \\
    {\it Top Right}: Model auto-correlation functions (ACFs) for the extracted light-curves, where the colours corresponds the the extraction regions in the SED. These have been calculated by considering 250 day chunks of the light-curves, and averaging the resulting ACF from each chunk, as current intensive monitoring data generally covers 200-300 day time-scales. It is clear that UVW2, U, and V are dominated by long term slow variability. HX on the other hand has a narrow peak due to the fast variability intrinsic to the hot corona, but is generally dominated by the slow variability propagating in from the disc. \rev{SX appears as something in between UVW2/U/V and HX, being clearly dominated by the long term, but with a weak reverberation signal giving a small peak.} \\
    {\it Bottom}: Model light-curves extracted from the time-dependent SEDs. The colours correspond to the extraction region in the top left panel. It is exceptionally clear that the model vastly over-predicts the variability in the X-ray.}
    \label{fig:modLC_hfrac1}
\end{figure*}

Now that we have a description of the variability in mass-accretion rate at each radial annulus we can calculate the time dependent SED, following the description in Section 2. Firstly, we generate realisations of $\mdot(t, r)$, following the previous section, for each radial annulus within the flow. This allows us to calculate the time-dependent emission from each annulus, which we then sum over to create the time-dependent SEDs. We note that calculating the SED at any given time is similar to calculating a single mean SED, with the difference being that we use a local $\mdot(t, r)$ at each radial annulus rather than a single $\mdot$ for the entire flow. 

For the standard disc region, the spectrum at each radius is directly given by $\mdot(t,r)$ as this sets the effective blackbody temperature. 
For the warm Compton region the optical depth and temperature are assumed constant so the only shape change is from the 
seed photon temperature which is set by reprocessing on the 
underlying passive disc to the 
effective blackbody disc temperature. 

The hot corona is more complex as the spectral index, $\Gamma_{h}$, and temperature, $\kTh$, change with the changing ratio between seed photon cooling, $\Lseed$ and gravitation heating, $\Ldiss$.
We set the power dissipated in annulii in the corona as $\Ldiss(t,r) \propto \mdot$, while $\Lseed(t, r)$ tracks the warm Compton power lagged by the light travel time. The difference in time dependence
of heating and cooling lead to changes in $\Gamma_{h}$ and $kT_{e,h}$
as described by equations \ref{eqn:gamma} and \ref{eqn:kth}.

Fig. \ref{fig:modLC_hfrac1} shows a resulting model realisation, using the underlying mean SED from Fig. \ref{fig:f9SED} and the variability parameters used for Fig. \ref{fig:example_mdot_powSpec_and_tseries}. For repeatability, we set the initial random seed to 1113, using the random number generator from {\sc numpy}. This seed will remain the same throughout the paper (unless otherwise stated), such that different model lightcurves can be directly compared, and also corresponds to the time-series in $\mdot$ shown in Fig.\ref{fig:example_mdot_powSpec_and_tseries}. Additionally, the radial resolution is set to $N_{\mathrm{dec}} = 500$. We
discard the first 200 days of our output time-series in order for the propagating fluctuations to fill the radial grid. 

Fig. \ref{fig:modLC_hfrac1}a shows 200 SEDs randomly selected from within the model time-series (orange lines) compared to the mean SED (black line). It is immediately obvious that the model has variability which is much larger in the EUV (warm disc peak) than in the UV/optical.
\rev{This is due to our assumption that the optical/UV/EUV emission is warm Comptonisation of seed photons from a disc. Each disc annulus has a spectrum below its peak  which is like the Rayleigh-Jeans tail of a blackbody at the disc effective temperature. This has monochromatic luminosity $L_{\nu} \propto \Mdot^{1/4}$, whereas around the peak the monochromatic luminosity goes more like the bolometric luminosity so $L_{\nu} \propto \Mdot$. Thus the disc variability around and above the peak (EUV/soft X-ray) is 
strongly enhanced relative to that in the Rayleigh-Jeans tail (optical/UV).}

\rev{The predicted variability is even stronger in the hard X-ray part of the spectrum. The warm disc peak produces a variable seed photon flux, which adds to the variability produced by the propagating fluctuations in modulating the slow variability in the hard X-ray corona. There is also faster variability generated in the corona itself.}

We quantify this energy dependence of the variabilty by extracting lightcurves in Swift UVOT bands 
(UVW2: magenta, U: green and V: orange) by multiplying the SEDs with each 
filter effective area curve. 
\rev{We similarly extract soft and hard X-ray light-curves from 0.3-1.5\,keV (hereafter SX: cyan) and 1.5-10\,keV (hereafter HX: blue) by multiplying the absorbed spectrum by the XRT effective area, \citet{Rapisarda16} (their equation A5).}

The resulting lightcurves are shown in the lower panel of Fig. \ref{fig:modLC_hfrac1}. The V/U/UVW2 bands show the typical amount of variability seen in Fairall 9 on timescales of $\sim 1$ year 
(\citealt{Hernandez20}, Edelson et al. (submitted)), 
\rev{but the model predicts HX variability that is much larger than observed in the data.}

\rev{The timescale of the HX variability is also not a good match to the observed data.  Fig. \ref{fig:modLC_hfrac1} (upper left) shows the autocorrelation function (ACF) of each lightcurve, and it is clear that while HX does have a narrow core, indicating its fast variability, the rest of its ACF is very similar to the other bands which are dominated by the disc. }

Thus the model gives reasonable amplitude (few 10s of percent over timescales of 6-12 months) variability in the optical/UV bands, but 
predicts too much large amplitude, slow 
variability 
in the corresponding HX lightcurve, compared to what is seen in the X-ray data from Fairall 9. 

\subsection{Suppressing the Variability Seen by the Corona}

\begin{figure*}
    \includegraphics[width=\textwidth]{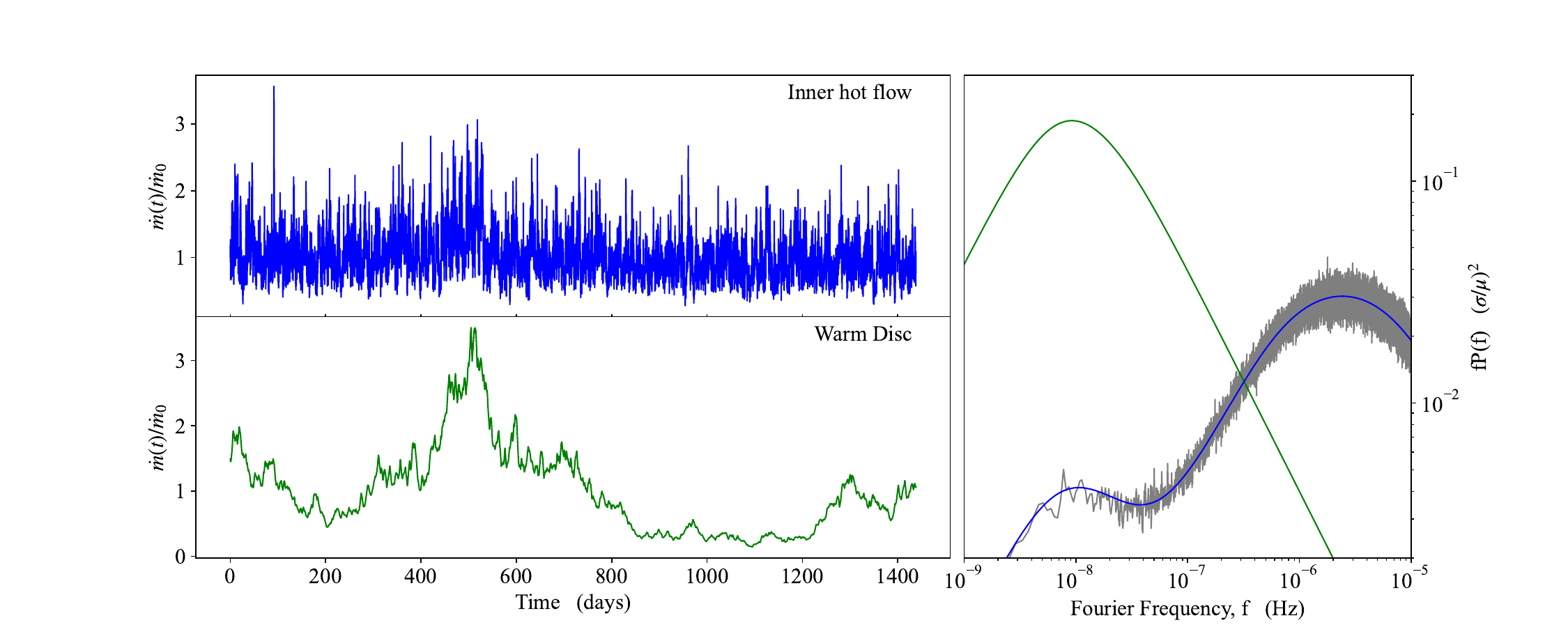}
    \caption{Same as Fig. \ref{fig:example_mdot_powSpec_and_tseries}, but now suppressing the disc variability seen by the hot corona by $f_{h}=0.02$. It is clear in the power-spectrum that the fluctuatios within the hot corona should now be dominated by rapid variability, with only a small contribution from the slow. Indeed, it can be seen in the time-series that the inner hot flow no longer contains a significant slow trend remesbling the warm; unlike the case in Fig. \ref{fig:example_mdot_powSpec_and_tseries}}
    \label{fig:mdot_pspec_hfrac02}
\end{figure*}

The overprediction of the X-ray variability is a surprise in the light of the results from the stellar mass black hole binaries. These 
strongly require that fluctuations propagate from the warm/turbulent disc into the corona, unhindered by strong viscous diffusion \citealt{Rapisarda16,Mahmoud19,Kawamura23}. Yet the standard disc equations predict that 
viscosity should spread out the fluctuations on the propagation timescale \citealt{Mushtukov18, Kawamura23}). The hot corona is likely in the regime where fluctuations can propagate in a wavelike manner at the sound crossing time \citep{Ingram16} i.e. have $\alpha \leq H/R$, but the warm/turbulent disc is at a much lower temperature, so has much lower $H/R$, hence is more likely to be in the regime where viscous diffusion dominates. 
The stellar mass black holes likely only generate variability on the inner edge of the truncated disc, so the fluctuations could 
could be generated by strong turbulence at this point rather than variability propagating through the disc itself. However, here we have considered the warm disc to be intrinsically variable across its entire radial range.  

Clearly though there is an issue in our assumptions 
\rev{as the model does not reproduce the observed variability. The two likely 
culprits are either our SED shape, or the assumption that all the disc variability is propagated into the X-ray corona in a lossless fashion. In the case of the SED shape, it is possible that the disc spectrum is described by a process that makes its luminosity proportional to $\Mdot$ everywhere, rather than giving enhanced variability around its peak.}
This would lead to UVW2 being a $\sim 1:1$ tracer of the intrinsic disc variability, eliminating the issue where 
the disc peak produces such strong seed photon variability which enhances the variability from propagated fluctuations. However, for the emission mechanisms generally associated with accretion discs (multi-colour black body, Comptonised black-body) this would imply that UVW2 must see the peak emission. The presence of a soft X-ray excess that appears to point back to the UV down-turn in the majority of AGN (e.g \citealt{Laor97, Porquet04, Gierlinski04}) would suggest an SED shape that links the EUV and soft X-ray emission to the disc. As UVW2 is generally below the UV down-turn, this feature becomes incompatible with UVW2 seeing the peak emission. Additionally, timing studies show that the He II line, often used as a proxy for the ionising EUV (e.g \citealt{Mathews87, Baskin13, Ferland20}, displays stronger variability than the optical continuum (e.g \citet{Homan23}). 
\rev{This makes it more likely that the EUV variability is indeed  larger than in UVW2, as predicted by the model}.

The other possibility then is that not all of the warm disc variability propagates into the hot corona. For the remainder of the paper we will allow the mass accretion rate fluctuations from the warm disc to be suppressed by some factor before 
variability is suppressed somehow before propagating into the hot corona. We stress, however, that the physical mechanisms that could cause this is not understood. Hence, the following should be treated as a phenomenological approach.

The simplest way to reduce the variability that propagates into the hot
corona is by setting $\sigma_{h}(\mdot, r_{k}) = f_{h} \sigma(\mdot, 
r_{k})$, where $\sigma(\mdot, r_{k})$ is the variance of $\mdot(t)$ at 
the radial annulus $r_{k}$; taken here to be the inner annulus of the 
warm region; $\sigma_{h}$ is the variance of this annulus as seen by the 
hot corona, and $f_{h}$ is a scaling fraction, with $0 \leq f_{h} \leq 
1$ (see e.g. \citealt{Mahmoud19}). The mass-accretion rate time series 
from the inner annulus of the warm region, $\mdot(r_{k}, t)$, as seen by 
the hot corona, $\mdot_{h}(r_{k}, t)$, is then:

\begin{equation}
    \mdot_{h}(r_{k}, t) = \big(\mdot(r_{k}, t) - \mdot_{0}\big) \sqrt{f_{h}} + \mdot_{0}
\end{equation}

This allows fluctuations to propagate without losses through the disc, but then only 
a fraction $f_{h}$ of these propagate 
into the hot corona. 
We stress that this suppression only affects the matter propagation, the soft seed photons for the corona are still modulated by the total disc variability.

We perform a run of the same model as in Fig. \ref{fig:modLC_hfrac1}, but setting $f_{h}=0.02$. This gives an X-ray time-series that is dominated by the fast variability, generated within the corona itself, but with a small contribution from a slow component propagating in from the disc and its slow seed photon variability. 

Fig. \ref{fig:mdot_pspec_hfrac02} (left) shows the time dependent of the mass accretion rates though the corona (upper) and warm disc (lower) in this version of the model, together with their power spectra (right). The warm disc is still the same as before 
(compare to Fig. \ref{fig:example_mdot_powSpec_and_tseries}), but the 
power spectrum of the mass accretion rate through the corona is now dominated by the 
fast variability, though it does still contain
a small bump at lower frequencies originating from the warm disc. 
Importantly, this slow variability present within the hot corona is still be correlated with the warm disc variability. Any resulting X-ray light-curve will then have a marginal correlation with the slowly variable UV/optical, giving rise to a disconnect driven by the strong fast variability within the hot corona.

\begin{figure*}
    \includegraphics[width=\textwidth]{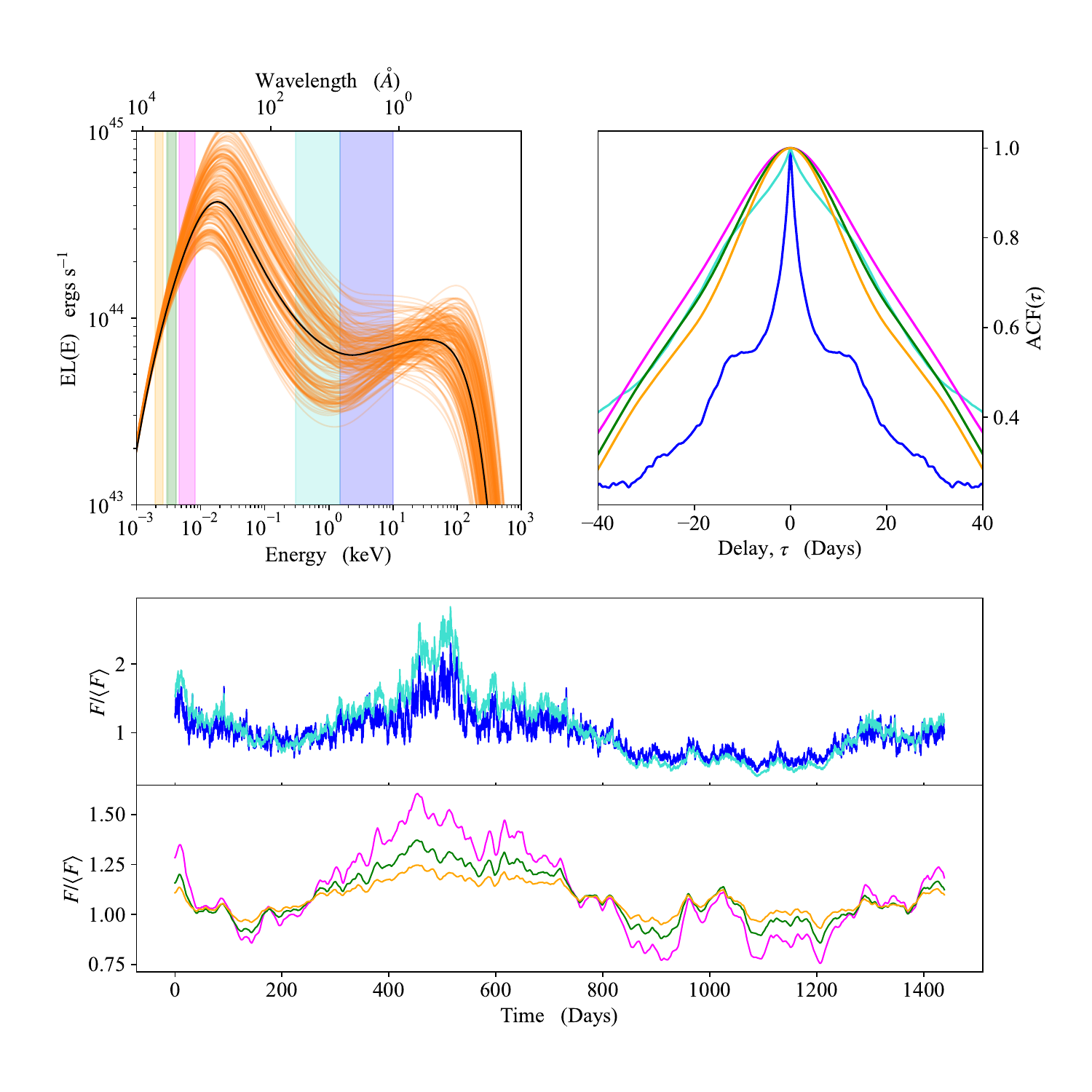}
    \caption{Example model output for a run considering the intrinsic variability only, using the $\mdot$ realisation from Fig. \ref{fig:mdot_pspec_hfrac02}.\\
    {\it Top Left}: The output SEDs. The solid black line shows the \rev{input SED calculated from the SED fit in section 2.1}, while the orange lines shows a sample of 200 SEDs randomly selected from the variable output. The shaded regions show the extraction regions used to generate light curves in HX (blue), \rev{SX (cyan)}, UVW2 (magenta), U (green) and (V) orange. For UVW2, U, and V the Swift-UVOT effective area curves were used to calculate the light-curves. \\
    {\it Top Right}: Model auto-correlation functions (ACFs) for the extracted light-curves, where the colours corresponds the the extraction regions in the SED, again calculated using 250 day chunks of the light-curves. It is clear that UVW2, U, and V are dominated by long term slow variability, whereas HX has a narrow core due to the rapid variability intrinsic to the hot corona and a broad base from the slow variability that propagates into the corona from the disc. \rev{Unlike HX, SX is not significantly changed after the suppression, being still clearly dominated by a slow component, due to the soft X-ray excess significantly contributing to this band-pass.}\\
    {\it Bottom}: Model light-curves extracted from the time-dependent SEDs. The colours correspond to the extraction region in the top left panel.}
    \label{fig:modLC_intrinsic}
\end{figure*}

\begin{figure*}
    \centering
    \includegraphics[width=\textwidth]{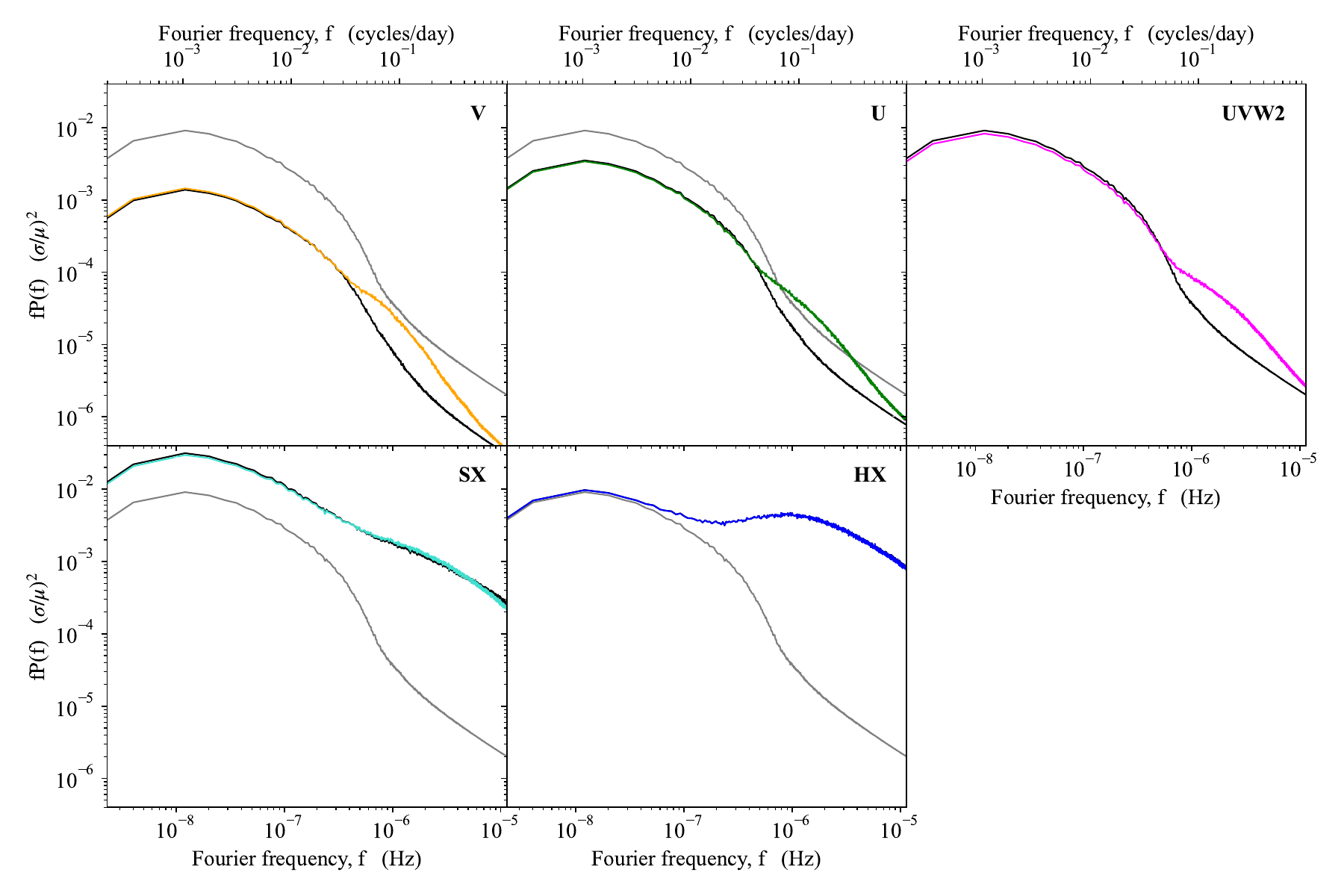}
    \caption{Model power-spectra for HX, \rev{SX}, UVW, U, and V band light-curves, for models considering intrinsic variability only (black lines), and intrinsic and disc reverberation (coloured lines). \rev{The grey lines in each panel show the intrinsic only power-spectrum for UVW2 as a comparison.} Due to the stochastic nature of the model, the power-spectra were created by calculating the averaged power-spectrum from 1000 $\mdot$ realisations of each model. Including the disc reverberation clearly adds a high frequency contribution to the total power, but has almost no impact on the slow variability or the total (integrated) power.}
    \label{fig:mod_powSpec}
\end{figure*}

We convert this new set of $\mdot(r,t)$ fluctuations into a time dependent SED. The model and resulting lightcurves are shown in Fig. \ref{fig:modLC_intrinsic}. 
\rev{As expected, there is a significant reduction in the variability of the X-ray portion of the SED
(compare the upper left panels of Fig. \ref{fig:modLC_intrinsic} and 
Fig. \ref{fig:modLC_hfrac1}), which gives a much better match to the properties of the data. The lower panel shows the lightcurves in the optical/UV are the same as before 
(compare with Fig. \ref{fig:modLC_hfrac1}), but that the scale of the HX lightcurve is now smaller. The model now also reproduces the disconnect in variability timescale, as shown by the ACFs (upper right panel of \ref{fig:modLC_intrinsic} and Fig. \ref{fig:modLC_hfrac1}). The warm disc optical/UV ACFs are the same as before, but now the HX ACF (blue) is dominated by the short timescale variability giving the narrow core,
with much lower correlation coefficient ($\sim 0.3$) for the long timescale wings.}

\rev{
The SX ACF (cyan) is much more like those from the optical/UV (orange/green/magenta). This is because the SX band-pass is dominated by the hottest part of the warm disc, so is dominated by this slow variability component, though its amplitude should be larger than in UVW2
(see the lightcurves in the lower panel). 
There is also a small contribution to the SX bandpass of the low energy emission of the hot corona. This fast variable component gives the small narrow core to the SX ACF. }

\section{Reverberation}

\subsection{X-ray illumination of the disc}

We now have a model that can describe the variability intrinsic to the flow. However, since a portion of the X-ray photons emitted by the corona are incident on the disc,  this imprints additional variability on to the UV/optical with an additional lag originating from the light-travel time (e.g \citealt{Blandford82, Welsh91, McHardy14, Edelson15}). We directly calculate the effect of this assuming the X-ryas thermalise at the local blackbody temperature, as first done by \citet{Gardner17}, and later by \citet{Mahmoud20, Mahmoud23}; \citetalias{Hagen23a}.

\begin{figure*}
    \includegraphics[width=\textwidth]{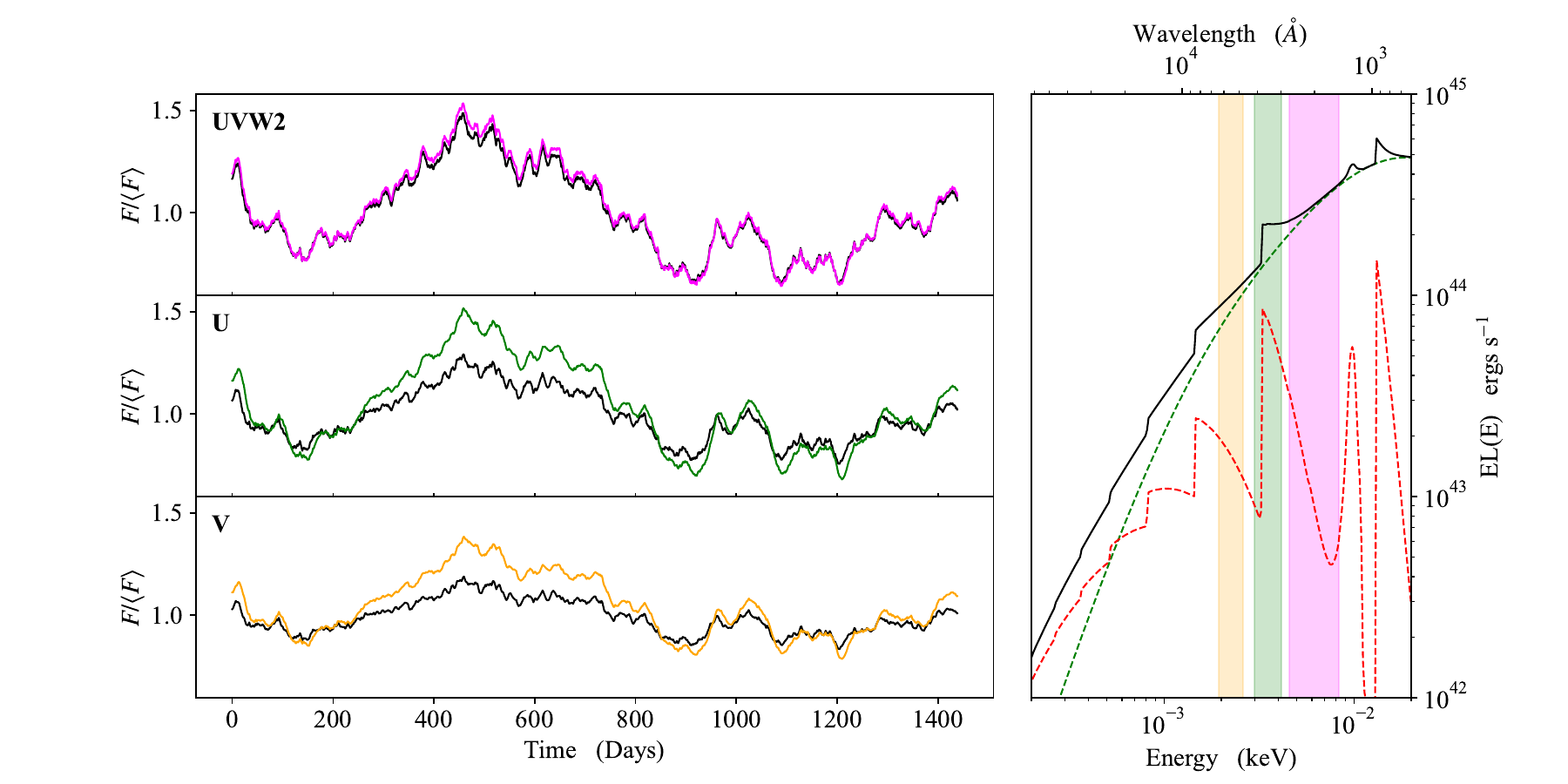}
    \caption{{\it Left:} Model light-curve, using the same $\mdot$ realisation as in Fig. \ref{fig:mdot_pspec_hfrac02}, calculated for a system with and without a wind (coloured and black lines respectively). Additionally, propagation and disc reverberation is included in both scenarios. We see a clear increase in response in both U and V bands due to the wind, since these have a stronger contribution from the free-bound continuum than UVW2, as seen in the right panel. \\
    {\it Right}: optical/UV SED. The dashed green line shows the intrinsic emission (i.e warm Compton component from the disc-like structure), while the dashed red line shows the free-bound continuum from the wind. The solid black line shows the total SED. The coloured panels show the extraction regions for UVW2, U, and V bands, where the colours correspond to the light-curves on the left. Note that we have subtracted out the line-emission from the free-bound component, for both clarity and computational efficiency.}
    \label{fig:wnd_vs_nownd_LC}
\end{figure*}

For details on calculating the re-processed variability we refer the reader to \citetalias{Hagen23a} (sections 2.2 through to 2.4). 
This imprints variability on a similar time-scales to the intrinsic X-ray variability, since the strongest response comes from close to the inner edge of the disc where the illuminating X-ray flux is also the strongest \citepalias{Hagen23a} and the lag the shortest
($\sim 0.1$\,days, light travel time to the warm disc), whereas the light travel time to the outer disc edge is of order $\sim 2$\,days. 
This puts some small fraction of the fast variability into the disc emission, but the effect of this is small. Fig.\ref{fig:mod_powSpec} shows the power spectra of the HX, \rev{SX, }UVW2, U and V bands, with the black lines showing the intrinsic variability, and the coloured lines including the disc reprocessed flux. This thermal disk reprocessing only makes a difference in the V, U and UVW2 bands, giving a small addition to their power at the highest frequencies, but the  overall effect is small as shown directly in \citetalias{Hagen23a}. 

Hence we also consider the effect of reprocessing from a larger scale wind in order to increase both the amplitude of reprocessed variability and the lag timescale, as required by the data \citepalias{Hagen23a}.

\subsection{Including a wind}

There is growing evidence for a large scale height wind on the inner edge of the BLR. This is seen directly in broad blue-shifted UV absorption lines which correlate with 'neutral' time variable X-ray absorption \citep{Kaastra14, Cappi16, Mehdipour16, Dehghanian19b, Kara21, Netzer22}, as well as in the lag spectra, where there is a prominent jump in the U band from diffuse Balmer continuum emission \citep{Korista01, Korista19, Cackett18, Lawther18, Chelouche19}. 

We use the same geometric wind model as in \citetalias{Hagen23a} \rev{(see their figure A1)}, i.e. 
a bi-conical outflow, launched at radius $r_{l}$ at an angle $\alpha_{l}$ with respect to the disc. The wind extends to a maximum radius and height, $r_{w, \mathrm{max}}$ and $h_{w, \mathrm{max}}$ \rev{(in cylindrical coordinates)}, such that the wind has a total covering fraction $\fcov$ as seen from the central source (i.e centred on the black hole). 
We take wind column to $N_{H} = 10^{23}$\,cm$^{-2}$ and 
ionisation state of $\log\xi\lesssim 0$ so that moderate Z elements still have a complete K shell (e.g. CIV), guided by
HST observations of other AGN where the wind is seen directly
\citep{Mehdipour16, Kara21}.
We assume the wind is launched from $r_{l}=800$ as this gives typical lags of 5-6 days as seen in Fairall 9 \citep{Hernandez20}, and we set the 
launch angle of $\alpha_{l}=60$\,deg and total covering fraction of $\fcov=0.3$ (e.g \citealt{Baskin18}). This small distance requires 
a high wind density of $n_e=10^{13}$~cm$^{-3}$ to match the low ionisation state. This also means that 
the emission is dominated by bound-free continuua rather than lines. 
We use the {\sc cloudy} radiative transfer code \citep{Ferland17} to calculate emission from a constant density slab of these parameters (unlike \citetalias{Hagen23a} where the emission was simply modelled as a blackbody), to give an overall wind reprocessed spectrum for the mean, minimum and maximum luminosity SEDs. 
For a near face on object, as is likely the case for Fairall 9, the observer is looking down the bi-cone of the wind, so only see the emission reflected off the wind surface. Hence, we only extract the reflected component from the {\sc cloudy} output, not the diffuse and transmitted fluxes which are seen by an observer 
looking through the wind. 

We divide the the wind surface into a spherical polar grid in $\theta$ and $\phi$, with spacing $d\cos(\theta)=0.01$ and $d\phi=0.01$~radians.  Each grid-point is located at a distance $r_{w}$ from the central source, with corresponding time-lag $\tau_{w}$, so sees the SED shape from this time. We interpolate between the minimum and maxium SED wind models to calcluate the appropriate lagged emission, and sum over the wind surface. We do not currently include the expected Doppler shift on the wind emission 
as the features are already broad. 

\begin{figure*}
    \centering
    \includegraphics[width=\textwidth]{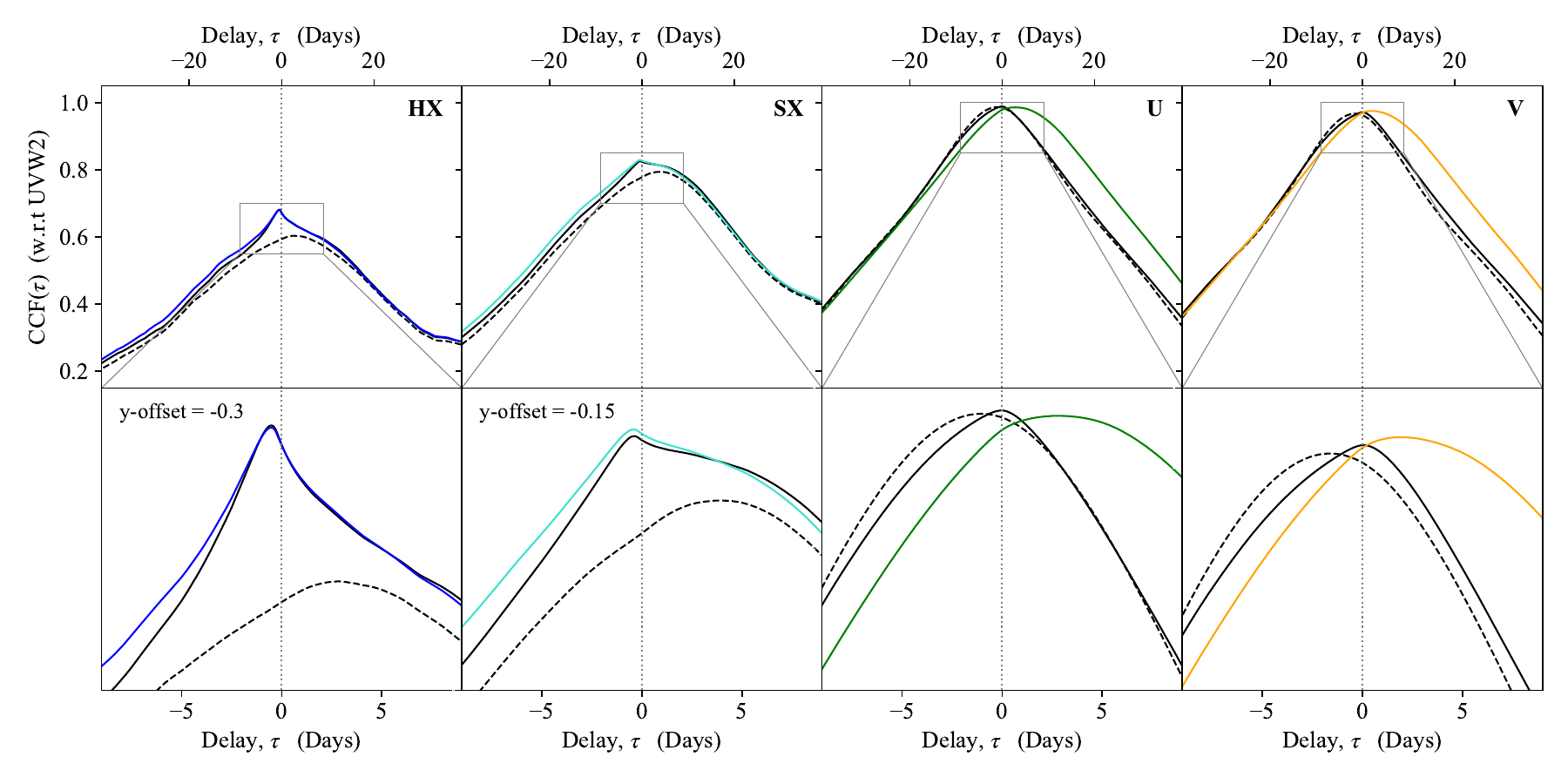}
    \caption{Cross correlation functions for HX, \rev{SX}, U, V band light-curves (segmented into chunks of 250 days each), with respect to UVW2, for the same $\mdot$ realisation as in Fig. \ref{fig:mdot_pspec_hfrac02} using only the intrinsic variability (dashed lines), intrinsic and disc reverberation (solid black line), and the full model including intrinsic variability, disc reverberation, and wind reprocessing (coloured solid lines). \rev{The top row shows the CCFs over the range $\tau \in [-39, 39]$\,days, while the bottom row shows a zoom-in of each CCF over the range $\tau_{\rm{zoom}} \in [-9, 9]$\,days. The zoom in plots cover the region within the grey boxes in the top panel. These boxes are the same size, with the SX and HX boxes offset in the y-direction with respect to U and V.}. The dotted vertical line indicates 0-lag. It is clear that the disconnect between the X-ray and UV/optical is driven by the intrinsic variability, while including disc and wind reverberation will shift the peak CCF from inward lags (i.e optical before UV/X-ray) to outward lags.}
    \label{fig:mod_ccfs}
\end{figure*}

Fig. \ref{fig:wnd_vs_nownd_LC} (left panel) shows the resultant lightcurves in UVW2, U and V together with the  UV/optical SED (right panel). The diffuse wind emission (red dashed line: right panel) is dominated by the recombination continuua (Lyman, Balmer, Pashen etc.) and Rayleigh scattering of the Lyman alpha line \citep{Korista01}. These make little difference to the UVW2 bandpass (right panel, magenta shaded), but the Balmer continuum makes a large contribution to the U band (green shaded), while Pashen affects the V (orange shaded), so the variability in the corresponding lightcurves (left panel) are clearly boosted in V and U. This additional variability is very similar in shape and time-scales to the intrinsic variability originating from the disc, as can be seen both by eye in Fig. \ref{fig:wnd_vs_nownd_LC} and in the cross-correlation functions in the next section (Fig. \ref{fig:mod_ccfs}). Unlike the black-body models of \citetalias{Hagen23a} the {\sc cloudy} models are responding to the UV/EUV emission, rather than the X-ray. This is because for a column-density of $10^{23}$\,cm$^{-2}$ the wind will be optically thin to X-rays above $\sim 3$\,keV, and so most of the X-ray flux is not re-processed. On the other hand, this column is very optically thick to the EUV emission. Hence it will respond to and re-process the energetically dominant, slowly variable, EUV component, giving an increase in the overall variability power on long time-scales.

\section{Lags from Simulated light-curves}
\label{sec:lags_from_sims}

\subsection{Lags from CCFs}

Fig. \ref{fig:mod_ccfs} shows the \rev{model} cross-correlation 
functions for Swift HX (1.5-10\,keV), 
\rev{SX (0.3-1.5\,keV)}, U, and V bands, calculated \rev{from our model light-curves chuncked into segments of 250\,days} following \citet{Gardner17}.
\rev{The upper panel shows the longer term cross-correlations ($\pm \sim 40$~days) while the lower panel zooms in on the shorter timescale behaviour ($\pm \sim 10$~days). The black dashed lines show the models which consider propagation only, the black solid lines disc
reverberation from variable X-ray illumination, while the coloured lines 
show results from the full model which also includes reverberation of the entire SED from the wind.} 

\rev{
We start by describing the intrinsic fluctuations i.e. the propagation only model (dashed black line).
The correlation normalisation of the X-ray lightcurves is significantly worse than those for U and V. This is as expected, and highlights the result
from the previous section that the UV-X-ray disconnect can be explained through propagating fluctuations. There are two 
distinct regions (disc/warm corona and hot corona) varying on intrinsically on different time scales, but linked through propagation. All the UV/optical lightcurves 
originate from the warm disc, so are highly correlated with each other, though there are signs of propagation in that the V band and U band lead UVW2 (peak at negative delay). Instead, the X-ray lightcurves are much less correlated with UVW2, as the X-ray variability is dominated by the additional uncorrelated fast variability stirred up the corona itself. We stress that it is not enough to simply have two distinct regions varying completely separately from one another, as  this would give a complete disconnect between the UV/optical and the X-ray, as the resulting light-curves would be incoherent. Instead, 
the data show correlations between the X-ray and UV/optical that are better than one would expect for two separate incoherent processes, but much worse than expected for a single process (e.g \citealt{Edelson19}). 
The direction of the propagated (correlated) variability goes from UVW2 to the X-ray corona, so the X-rays lag UVW2 (peak at positive delay).
}

\rev{
The black solid lines in Fig. \ref{fig:mod_ccfs} show the results after including X-ray reverberation from the disc.
This adds a small peak at short timescales in all the CCFs as it produces a fast variable response in the warm disc 
which lags behind the X-ray lightcurves. Thus the fast variability in the X-ray bandpass leads its reprocessed signal in UVW2, giving a peak in the CCF lag at $\sim -0.5$~days for HX. Conversely, for all the warm disc bands (UVW2 as well as U and V) the 
reprocessed X-ray variability is all dominated by the component produced on the inner disc edge \citepalias{Hagen23a}
so the U and V CCFs now peak at zero lag with respect to UVW2. 
}

\rev{
The coloured lines show the CCFs from the full models, including the wind reverberation. 
This makes a large difference to the U and V band lightcurves, but has very little contribution to 
UVW2 and the X-ray bands (see Fig.\ref{fig:wnd_vs_nownd_LC}). The wind imprints the 
warm disc variability with light travel time lag via the Balmer (U band) and Pashen (V band) diffuse continuua,
moving the CCF peak from zero to a progressively longer lags in U and V behind UVW2. The increase in lag at V compared to U is not from the reprocessing picking out structures at larger distance, but is instead from the increased 
fraction of the constant lagged wind at V compared to U. 
}

\subsection{UV/optical lag spectrum}
\label{subsec:lag_spec}

So far we have only compared light-curves extracted for broad-band filters (in particular Swift-UVOT filters). However, our model predicts the full variable SED, and as such allows us to extract light-curves at the spectral resolution of the model ($d\log (E/\mathrm{keV}) = 1/125$). This then allows us to predict the lag-spectrum one would expect from each model, by comparing the model light-curve in each energy bin to a single reference light-curve. In this case we pick our reference as $\lambda_{\mathrm{ref}} = 1928$\,\AA, as this is the centre of the Swift-UVOT UVW2 bandpass, and as such gives the cleanest comparison to current intensive monitoring campaigns. 

For each light-curve extracted from each energy (wavelength) bin in our variable SED, we calculate the cross-correlation function with respect to the light-curve in the energy bin that covers $1928$\,\AA, \rev{again with the light-curves chunked into segments of 250\,days}. We then use these CCFs to extract an estimate for the model lag as a function of energy (wavelength), defined as the lag that corresponds to the maximal correlation coefficient. 
We measure this centroid lag following the method used in the data analysis papers (e.g \citealt{Peterson98, Edelson19}); \rev{that is, restricting the centroid fit to just the region of the CCF with $ R > 0.8 R_{\mathrm{max}}$, where $R$ is the correlation coefficient $CCF(\tau)$ and $R_{\mathrm{max}}$ is the maximum measured value of $R$.}

\begin{figure}
    \centering
    \includegraphics[width=\columnwidth]{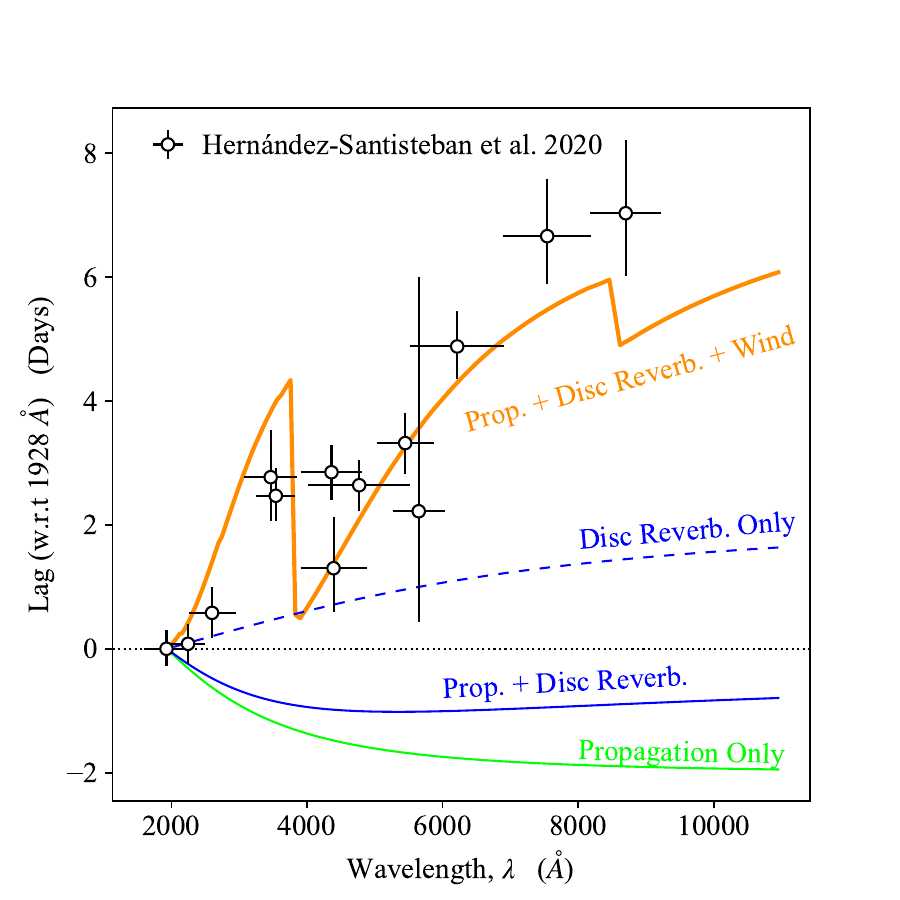}
    \caption{The data-point show the lags measured for Fairall 9 from \citet{Hernandez20}. The lines are lags from the models taken from the centroid of the CCF of the lightcurve at wavelength $\lambda$ with respect to 1928\,\AA. The solid lime green line shows results only including intrinsic propagation model. This gives negative (longer wavelengths lead shorter wavelengths) lags as the fluctuations start in the outer disc and propagate inwards. Including X-ray reverberation from the disc gives the solid blue line. The X-ray reverberation is too small to cancel the negative lag from propagation, so the lightcurves including both propagation and disc reverberation still have longer wavelengths leading. The dashed blue line shows the effect of X-ray reverberation from the disc alone without the intrinsic propagation, showing how these predicted reverberation lags are smaller than measured from the data. Finally, the orange solid line shows the lags from the full model, including UV (and X-ray) reverberation from a wind (as well as propagation and disc reverberation). Here we see a clear increase in lag with wavelength due to the increased contribution of the wind at longer wavelengths, with clear features corresponding to the Balmer and Paschen continua, Fig. \ref{fig:wnd_vs_nownd_LC}.}
    \label{fig:lagplot}
\end{figure}

We plot the resulting optical/UV lag relative to UVW2 in 
Fig. \ref{fig:lagplot}, except now we show lags at full model resolution rather than just extract over the observed photometric 
bandpass. For the propagation only model (green line) we see a strong negative lag, increasing to lower energy (higher wavelength), as expected for propagating fluctuations. An interesting point to note here is the measured lag 
of ($\sim -2$\,days) is considerably shorter than the model propagation time of $\sim 20$\,days
between the inner and outer disc radii 
(chosen for computational efficiency rather than physical expectations). This is due to blackbody emission being broad, so a single wavelength does not just contain emission from a single radius. 
Hence the observed time-delay will instead be more representative of the propagation time between the flux weighted radii for each energy, which will naturally be smaller than the difference between the outer and inner edge of the disc. 

When we include disc reverberation (blue line in Fig. \ref{fig:lagplot}) we see a reduction in the negative lag, not a switch to positive. This is interesting, as it shows that X-ray reverberation from the disc reverberation does not contribute enough signal to the light-curves to overcome the negative propagation lag. We also show (blue dashed line) the reverberation lag alone, without any underlying propagation through the disc. This clearly shows that X-ray reverberation in the disc gives a lag which is a few times smaller than that measured, and it is then clear that the propagation plus X-ray disc reverberation lag prediction is approximately the sum of the negative propagation lag and the positive disc reverberation lag. 

Including reverberation from the wind is the key to matching the data 
(orange line in Fig. \ref{fig:lagplot}), compared to the data points from \citet{Hernandez20}. 
The lag-wavelength relation calculated for the full model (propagation, disc reverberation, and wind reverberation) follows the data remarkably well, especially as this is not a best fit. The model parameters (generator and propagation timescales, wind size scale) were simply examples given to roughly match some of the observed properties of the data, yet the model has roughly both the correct shape and normalisation.

\begin{figure*}
    \includegraphics[width=\textwidth]{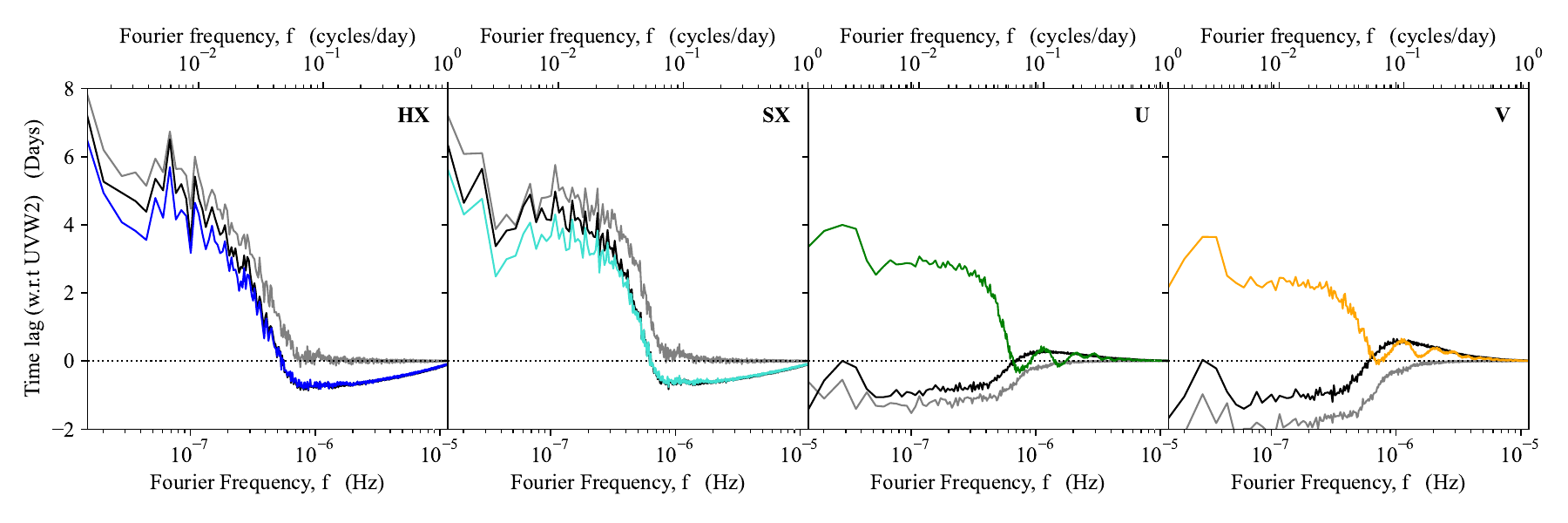}
    \caption{\rev{
    Time lags between UVW2 and HX (1.5-10\,keV), SX (0.3-1.5\,keV), U, and V as a function of Fourier frequency. These are defined such that a positive lag implies the relevant band follows (i.e lags) UVW2, while a negative lag implies UVW2 lags the relevant band. The coloured lines show the lags for the full model including propagation, disc reverberation, and wind reverberation; the black lines show the lags for a simulation considering only propagation {\it and} disc reverberation; while the grey lines show a model using {\it only} propagating fluctuations. The dotted black horizontal line show 0 lag. These have all been calculated by averaging over the Fourier lags calculated from 1000 simulation realisations. It is clear that disc reverberation only affects the high frequency (short time-scale) component, whereas the wind reverberation affects the variability on all time-scales due to the wind responding to the full EUV variability, which includes the intrinsic long-time scale variations. We note that the oscillatory features in the U and V band lags at high frequencies occur due to phase-wrapping between grid-points on the wind surface (i.e a resolution issue).
    }}
    \label{fig:fourier_lag}
\end{figure*}

The wind reprocesses the entire SED, but this is dominated by the EUV band, so the total light-curve in each band is a superposition of the light-curve emitted by the disc, and that reprocessed by the wind, which will be near identical to that of the disc but with some lag and additional smoothing on short time-scales. 
The measured lag rises with wavelength, not because the wind lag is intrinsically longer, but because the wind contributes more to the spectrum at longer wavelengths (see Fig. \ref{fig:wnd_vs_nownd_LC}).
The spectrum at each wavelength has a contribution from the intrinsic disc, which is highly correlated with the small (negative) lag at each wavelength, and the wind reverberation, which is lagged by a constant (positive) value $\tau_{\mathrm{wind}}$. The measured lag is then the flux weighted lag from each component, so $\tau\approx (F_{\mathrm{disk}}(\lambda)\tau_{\mathrm{disk}}(\lambda) + F_{\mathrm{wind}}(\lambda)\tau_{\mathrm{wind}})/F_{\mathrm{tot}}$, where $\tau_{\mathrm{wind}}$ is constant with wavelength. Only when the wind is the only component in the spectrum does the lag tend to $\tau_{\mathrm{wind}}$, otherwise it is {\em diluted} \citep{Uttley14} by the intrinsic disc emission which is in both the lightcurves (reference band at $1928$\,\AA\,\, which has almost no wind contribution and at $\lambda$). This gives the characteristic wind lag shape, with lags increasing at longer wavelengths, with sharp features from the Balmer and Paschen edges superimposed, and our example model parameters give a fairly good quantatative match to the observed lags despite not being fit to the data. 

\rev{
\subsection{The UV-Xray connection seen through Fourier Lags}
}
\label{sec:fourier_lags}


\rev{
Our model is clearly able to re-produce the phenomenology seen in IBRM campaigns, analysed using standard CCF techniques. However, CCFs are not ideal for disentangling multiple variability signals on different time-scales, while a key feature of our model is the presence of multiple signals varying on distinctly different time-scales, with different lags! A Fourier analysis, which calculates the different lags from different variability timescale components, is considerably better suited for this. 
Fourier resolved lags were first applied to the the higher signal-to-noise X-ray lightcurves from black hole binaries, where
they showed a complex pattern of lags where hard X-rays lagged softer ones, by an amount which depended on variability timescale. This led to the development of propagating fluctuation models, including disc reverberation to give an additional short lag for fast variability from a reflected component (e.g. the review by \citealt{Uttley14}), 
The much higher masses of AGN means that single X-ray observations are mainly sensitive only to the fast timescales of 
lagged reflection from disc reverberation, so X-ray AGN studies have focussed on this component 
\citep{DeMarco13, Zoghbi13, Cackett22}. Here we apply these techniques to illustrate how they can disentangle 
the multiple variability components in the model. 
}

\rev{
Fig. \ref{fig:fourier_lag} shows the time-lags with respect to the UVW2 bandpass as a function of Fourier frequency for HX (1.5-10\,keV), SX (0.3-1.5\,keV), U, and V band light-curves, calculated following \citet{Uttley14}. These are defined such that a positive lag implies UVW2 is leading, while a negative lag implies UVW2 is following. Similarly to  the CCFs in Fig. \ref{fig:mod_ccfs} we show lag curves for simulation runs considering {\it propagation only} (grey), {\it propagation and disc reverberation} (black), and {\it propagation and disc reverberation and wind reverberation} (coloured). 
}

\rev{
The propagation only model (grey) has the slow fluctuations starting first in the optical, then propagating into the U and UVW2 bands (so V and U lead UVW2) then into SX and HX (so these lag UVW2). Fast variability ($f>10^{-6}$~Hz) is only produced in the X-ray corona, but there is no correlation of this with the warm disc emission so the lag drops to zero in all bands. 
Including disc reverberation gives a fast variable component in UVW2 and U and V 
correlated with and lagged behind the X-rays. Thus the fast variable X-rays go first, then there is a 
response in UVW2 lagged on the light travel time to the disc of around $0.5$~days. The inner disc reprocessing 
dominates the entire warm disc \citepalias{Hagen23a}, so is also included in U and even V,  
diluting the lag from that expected from the outer disc size scales, but still giving a correlated fast variability signal
with the opposite sign of lag to that of the slow variability. 
}

\rev{
Including the wind makes very little difference to HX and SX as neither of these, nor the comparison lightcurve of UVW2, have any significant wind component. Instead, the wind reverberation makes a dramatic change to the slow variability in U and V, as the wind reverberation signal is mainly due to the slow variable EUV, but with a clear long lag from light travel time to the wind. The fast timescale lags from disc reprocessing are almost unaffected (the oscillatory behaviour at high frequencies is an artefact originating from the grid resolution on the wind surface leading to a phase-wrapping effect).
}
\\

\section{Conclusions}

The intensive broadband monitoring campaigns on AGN give simultaneous lightcurves from optical to X-ray energies. These contain much more information than simply the lag between any two bands. The amount of lagged signal is also important as a diagnostic
of the geometry, as is the timescale of variability. A reverberation origin for the UV variability predicts a lightcurve which is a lagged and smoothed version of the driving X-ray lightcurve, with lag timescale similar to the smoothing timescale. This is in clear conflict with the data, where the UV lags behind the X-rays by $\sim 1$~day, while it is smoothed on timescales of $\sim 20$~days. This is the main issue with X-ray reverberation models, not that the lag timescale is a factor $\sim 2$ bigger than predicted but that the X-ray and UV lightcurves are so different.  

Instead, we develop a full spectral-timing model to use all of the information in these intensive broadband monitoring datasets. We use the truncated (warm Compton) disc/hot inner flow geometry which successfully matches the SED in Fairall 9 (and other AGN with $L/L_{\mathrm{Edd}}\sim 0.05-0.2$). We assume that there are intrinsic fluctuations stirred up in the disc, which propagate down and modulate the much faster fluctuations stirred up in the hot flow. This produces intrinsic UV variability which is
much slower than the intrinsic X-ray variability, with the slow
UV variability correlated with and leading the slow component of the X-ray variability. The fast X-ray variability has no intrinsic correlation with the UV as it is stirred up only in the hot flow. Its reverberation on the truncated disc produces a very small amplitude, fast variable, lagged signal in the UV, but the UV variability is dominated by the intrinsic slow fluctuations. Instead, in our model the majority of the lag seen in the optical/UV arises from the intrinsic slow variable UV/EUV reverberating off a wind on the inner edge of the BLR. This gives a much larger amplitude signal than X-ray reverberation from the disc firstly as the EUV is where the SED peaks, so there is more intrinsic flux than in the X-rays, and secondly because a vertically extended wind intercepts much more flux from a central source than a flat disc. The wind reverberation signal is a lagged and smoothed version of the UV/EUV, but these are already intrinsically variable only on long timescales (unlike the X-rays), so there is no longer a mismatch between the lag and smoothing timescale. Perhaps the most un-intuitive aspect is that the increasing lag as a function of wavelength is not produced by increasing the scale of the reverberating structure at longer wavelengths. Instead, in this model it is produced by the increasing fraction of a fixed size scale reverberation signal at longer wavelengths, as bound-free continuum has a redder spectrum than the warm disc. 

Thus the  model succeeds in qualitatively explaining all the puzzling features of the Fairall 9 dataset, but it does have multiple free parameters even after the disc/hot flow radius is set by the SED. There is the intrinsic variability timescale as a function of radius, the propagation timescale as a function of radius, and the size scale of the wind. These were not derived from a fit to the data, but simply chosen from order of magnitude arguments about what was needed. 

The model predicts that the warm Compton disc has both SED and variability peaking in the EUV.
There is some evidence for increased variability in the EUV, e.g from He II photo-ionised line (e.g \citealt{Homan23}). However, this overpredicts the observed X-ray variability if all of the warm disc fluctuations propagate into the X-ray corona (unlike the BHB, see \citealt{Kawamura23}). 

\rev{
The model was tailored for moderate Eddington ratio AGN ($\LLedd \sim 0.05-0.2$), where there is a significant disc-like continuum present. 
It is unlikely to work for much 
lower Eddington ratio AGN ($\LLedd < 0.02$) where there is increasing evidence that the accretion structure changes significantly with the collapse of the warm disc component  
(e.g \citealt{Done07, Noda18}; Hagen et al. (submitted)). The fast coronal variability should still be present in these cases, 
however the nature of any remaining optical/UV variability is not at all clear, though it may be simply re-processing off some more distant material such as the BLR 
(e.g NGC 4151, \citealt{Edelson17, Mahmoud20}). 
}

\rev{
Our model also assumes a line of sight which does not intercept the wind. Similar AGN viewed at higher inclination 
angles would see additional variability from any wind variability, as seen in the STORM campaigns (e.g \citealt{Kara21, Partington23, Homayouni24}). This would significantly complicate the results, but would also allow the physical mechanism
producing the wind variability to be explored, which would help reveal its origin. 
}

\rev{
There are other more limited models that can explain some of the observed variability properties in AGN, e.g using stochastic temperature fluctuations within the disc \citep{Cai18, Cai20, Neustadt24} to give the smooth and slow UV variability, or a rapidly variable coronal height to cancel out the fast variations in the reprocessed UV signal \citep{Kammoun24}. However, our model provides a physical mechanism to both explain and predict the variability, based on the same mechanisms that are known to explain the variability in the better studied black hole binaries. In particular, it makes a prediction 
that there should be a switch between the UV {\it leading} the soft and hard X-rays for slow variability due to propagation down through the flow, and the UV {\it lagging} the hard X-ray variability for fast variability due to reverberation. This is best seen in a Fourier resolved analysis, and some of the intensive broadband monitoring campaign data is now sufficient to test this.
}

In summary, our model  can quantitatively match the observed lags between all the bands \rev{for unobscured, moderate Eddington ratio AGN}, and match the observed variability timescales (auto-correlation function widths) and amplitudes of variability in each band, particularly the disconnect between the disc dominated bands (optical/UV) and X-ray bands.
These successes
highlight the importance of its fundamental assumption, which is that the (warm Compton) disc is intrinsically variable. This requires that the internal disc structure is very different to the Shakura-Sunyaev prediction.
Understanding this would lead to breakthrough in our understanding of the 
energy generating structure in AGN. 

\section*{Acknowledgements}

We thank Juan Hern\'{a}ndez-Santisteban and the entire IBRM collaboration for 
all their efforts in organising these campaigns
and extracting the data, motivating this study.

\rev{We also thank the referee, Adam Ingram, for helpful comments, which improved the manuscript, especially for the suggestion to show the Fourier resolved analysis.}

SH  acknowledges support from the Science and Technology Facilities Council (STFC) through the studentship grant ST/V506643/1.  CD acknowledges support from STFC through grant ST/T000244/1.

This work used the DiRAC@Durham facility managed by the Institute for Computational Cosmology on behalf of the STFC DiRAC HPC Facility (www.dirac.ac.uk). The equipment was funded by BEIS capital funding via STFC capital grants ST/K00042X/1, ST/P002293/1, ST/R002371/1 and ST/S002502/1, Durham University and STFC operations grant ST/R000832/1. DiRAC is part of the National e-Infrastructure.

This work made use of the following {\sc python} modules: {\sc scipy} \citep{Virtanen20}, {\sc numpy} \citep{Harris20}, and {\sc astropy} \citep{Astropy13, Astropy18, Astropy22}. Additionally, all plots were made using {\sc matplotlib} \citep{Hunter07}.

\section*{Data Availability}

No new data were produced during this study. The SED in Fig. \ref{fig:f9SED} used archival data, available from HEASARC (\url{https://heasarc.gsfc.nasa.gov/cgi-bin/W3Browse/w3browse.pl}). The model code developed throughout this study is available via the corresponding authors GitHub (\url{https://github.com/scotthgn/AGNvar2}).



\bibliographystyle{mnras}
\bibliography{Refs} 




\appendix

\section{Modelling the wind variability}
\label{app:wind_mod}

To model the wind variability we start by considering a bi-conical outflow launched from radius $r_{l}$ at angle $\alpha_{l}$ with respect to the disc. The wind subtends a solid angle $\Omega = 4\pi \fcov$ as seen from the central black hole, where $\fcov$ is the covering fraction. This geometry is identical to that used in \citetalias{Hagen23a}. Here we provide details on the main calculations of the wind emission, but refer the reader to Appendix A in \citetalias{Hagen23a} for details on the geometry

Unlike \citetalias{Hagen23a} we use {\sc cloudy} (v.17.01 \citealt{Ferland17}) to calculate the emission from the wind. As we are mainly interested in the response of the free-bound emission to the ionising continuum we use a reatively simple {\sc cloudy} model, defined by the Hydrogen column desnity, $N_{H}$, Hydrogen number density, $n_{H}$, covering fraction, $\fcov$, and SED shape. The strength of the reflected free-bound continuum depends on the intensity of the ionising continuum at the wind, and so the distance from the illuminating source. The wind clearly subtends a range of radii from the central region. However, for simplicity we only calculate a single {\sc cloudy} run for the wind, at a distance set in the middle between the launch radius and maximal radial extent (defined through $\fcov$ - see \citetalias{Hagen23a}). This will give of the order correct emission power, and significantly reduces the computational cost when evaluating the time-dependent version later.

The wind is defined as launching from both sides of the disc, however the observer will only see the emission from the side facing the observer. Hence, for the total observed free-bound wind emission, $L_{\mathrm{wind}}$, we only extract half the total reflected emission given by {\sc cloudy}, such that:

\begin{equation}
    L_{\mathrm{wind}}(E) = \epsilon_{\mathrm{wind}}(E) 2\pi\fcov
\end{equation}

where $\epsilon_{\mathrm{wind}}(E)$ is the emissivity of the wind surface calculated by {\sc cloudy}.

We stress here that we use the reflected emission, as for our geometry the observer would be looking down the wind funnel, and therefore only see emission from the side of the wind facing the illuminating source, not the transmitted or diffuse components that are emitted from the opposite side (as defined by {\sc cloudy}). We also note that this assumes the wind emission is isotropic (within the lines of sight that will see the wind face), as you can expect from a diffuse medium.

To calculate the variability of the wind we also need to consider the light-travel time to different section of the wind. Hence, we start by dividing the wind into a polar grid in $\cos(\theta)$ and $\phi$, where $\phi$ is the azimuthal angle in the x-y plane and $\theta$ is the polar angle  measured from the z-axis. The grid is linearly space between $0$ and $2\pi$ with spacing $d\phi = 0.01$, and $0$ and $\cos(\theta_{m})$ with spacing $d\cos(\theta_{m}) = 0.01$, for $\phi$ and $\theta$ respectively. Here $\theta_{m}$ is the maximal polar angle of the wind, defined in terms of the covering fraction such that $\cos(\theta_{m})=\fcov$. As each grid-point is defined in terms of $\phi$ and $\theta$, the solid angle of each grid-point (as seen by the central source) is $d \Omega_{\mathrm{grid}} = d\cos(\theta) d\phi$. This solid angle is important, as we use it later to calculate the relative contribution from each grid-point when calculating the time-dependent emission.

The light-travel time to a grid-point on the wind surface is \citepalias{Hagen23a}:

\begin{multline}
    \tau_{w}(\theta, \phi) = \\
    \frac{R_{G}}{c} \left[ \sqrt{r_{w}^{2}(\theta) + h_{w}^{2}(\theta)}  - h_{w}(\theta)\cos(i) - r_{w}(\theta)\sin(i)\cos(\phi) \right]
    \label{eqn:tau_wnd}
\end{multline}

where $\rwind$ and $\hwind$ are the radius (in the x-y plane) and the height of the wind grid-point respectively, and are given by:

\begin{equation}
    \rwind(\theta) = \frac{r_{l}\tan(\alpha_{l})}{\tan(\alpha_{l}) - \tan(\pi/2 - \theta)}
\end{equation}

\begin{equation}
    \hwind(\theta) = \rwind(\theta) \tan(\pi/2 - \theta)
\end{equation}

We note here that Eqn. \ref{eqn:tau_wnd} is the delay assuming the distance from the central black-hole, whereas the equation in \citetalias{Hagen23a} also includes a term for the height of the corona above the black hole (assuming a lamppost geometry).

\begin{figure*}
    \centering
    \includegraphics[width=\textwidth]{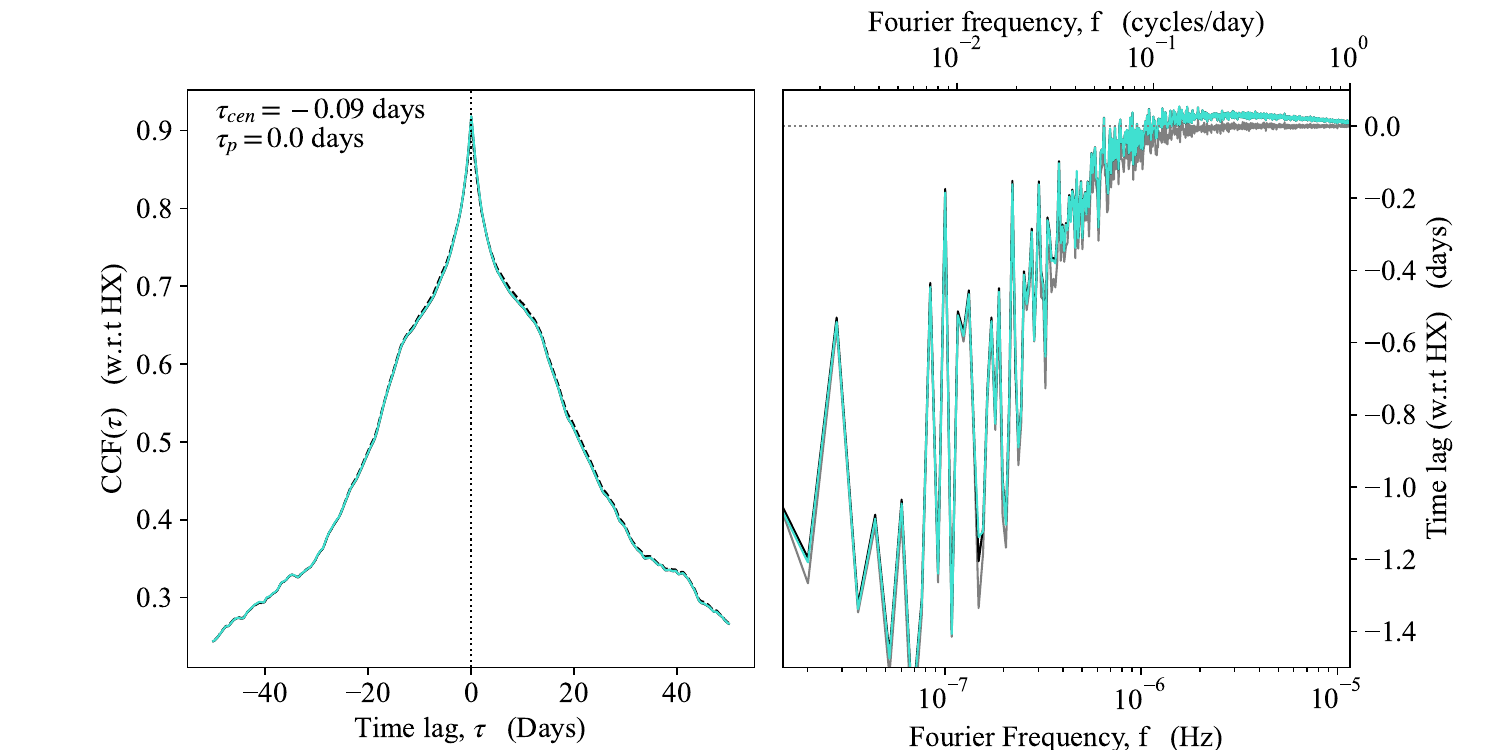}
    \caption{\rev{{\it Left:} Cross-correlation of SX with respect to HX light-curves. Here the peak is clearly dominated by the fast X-ray component, present in both bands, giving $0$ peak lag, while the slow variability from the soft X-ray excess (disc-like component) gives a marginal negative lag (i.e SX comes before HX) in the centroid lag. \\
    {\it Right:} Fourier lags between SX (0.3-1.5\,keV) and HX (1.5-10\,keV), defined such that a negative lag implies SX comes first followed by HX, and vice-versa for a positive lag. The magenta line shows the lags for the full model ({\it propagation, disc reverberation, and wind reverberation}), the black line shows the case for a model considering only propagation {\it and} disc reverberation, while the grey line shows the case for {\it only} propagation. It is clear that the wind makes no difference to the SX-HX relation, as the X-rays simply pass through this. Including disc reverberation gives a very small positive lag at the highest frequencies, whereas the low remain more or less unaffected, entirely dominated by propagation.
    }}
    \label{fig:sx_hx_ccf_frlag}
\end{figure*}

To calculate the wind emission at a given time we now assume the response to changes in the continuum are linear, bit not necessarily 1:1. This simplifies the problem significantly, as we now only run two {\sc cloudy} models. One for the minimum and another for the maximum luminosity values within the SED; $L_{\mathrm{min}}(E)$ and $L_{\mathrm{max}}(E)$. The two {\sc cloudy} models then give the minimum and maximum emissivity values of the free-bound wind emission; $\epsilon_{\mathrm{wind}, \mathrm{min}}(E)$ and $\epsilon_{\mathrm{wind}, \mathrm{max}}(E)$. The wind emission for any given intrinsic SED at any given time within our time-series (and hence within $L_{\mathrm{min}}(E)$ and $L_{\mathrm{max}}(E)$) is then given by:

\begin{equation}
    \epsilon_{\mathrm{wind}}(E) = f(E) \epsilon_{\mathrm{wind}, {min}}(E) + (1-f(E)) \epsilon_{\mathrm{wind}, {max}}(E)
\end{equation}

where $f(E)$ is an energy dependent interpolation factor given by:

\begin{equation}
    f(E) = \frac{L(E) - L_{\mathrm{max}}(E)}{L_{\mathrm{min}}(E) - L_{\mathrm{max}}(E)}
\end{equation}

where $L(E)$ is the intrinsic SED seen by the wind. We can make this time dependent by including the time-delay to a point on the wind by writing $L(E)$ as $L(E, t - \tau(\theta, \phi))$, such that $f(E)$ becomes $f(E, t, \theta, \phi)$. Of course, we need to take into account the size of a grid-point relative to the total wind area when calculating the time-dependent emission from each grid. For this we simply note that integrating $d\Omega_{\mathrm{grid}} = d\cos(\theta) d\phi$ over the wind surface will give $\Omega_{\mathrm{wind}}/2 = 2\pi \fcov$. Hence, the luminosity of a given grid point at time $t$ is simply 

\begin{equation}
L_{\mathrm{wind}, \mathrm{grid}}(E, t, \theta, \phi) = \epsilon_{\mathrm{wind}}(E, t, \theta, \phi) d\cos(\theta) d\phi
\end{equation}

and hence the total time-dependent wind emission is simply a sum over all grid points, such that:

\begin{equation}
    L_{\mathrm{wind}}(E, t) = \sum_{\cos(\theta)=0}^{\cos(\theta_{m})} \sum_{\phi=0}^{2\pi} L_{\mathrm{wind}, \mathrm{grid}} (E, t, \theta, \phi)
\end{equation}

This is of course just for the free-bound wind component. The total variable SED is then a sum of the intrinsic and wind components, such that $L_{\mathrm{tot}}(E, t) = L(E, t) + L_{\mathrm{wind}}(E, t)$.

\rev{
\section{The soft to hard X-ray connection}
}

\rev{
The paper focuses on the broadband continuum lags from optical, through UV to soft and hard X-rays. 
We generally show results of lags with respect to UVW2, as in generally done in the IBRM campaigns. However, 
our model covers the full SED, so gives predictions for X-ray lags as well. We repeat the approach in Section 5 
to calculate the CCF and Fourier resolved lags, but with HX as the reference lightcurve rather than UVW2. 
}

\rev{
Fig. \ref{fig:sx_hx_ccf_frlag}a shows the CCF of the SX band lightcurve with respect to HX
for the propagation only model (black dashed line), plus disk reverberation (black solid line) and full model 
including the wind (coloured). Wind reverberation makes very little impact as the wind emission has very little contribution in 
either X-ray band. Similarly, while the disc reverberation does contribute to the SX lightcurve, this is 
a very small effect so is almost indistinguishable from the intrinsic (propagation only) lightcurves
(see also  Fig.\ref{fig:mod_powSpec} so these different model lines are hard to separate. 
The main effect is that the SX lightcurve is a mix of the 
soft X-ray excess which originates from the innermost radii of the (slow variable) 
optically thick geometrically thin disc-like structure, and the low energy extent of the (fast variable) 
coronal emission. The HX component by contrast is purely the coronal emission, which has the propagated variability from the 
warm disc, modulating its own additional fast variability. 
Thus the SX-HX CCF has a perfectly correlated fast core from the same corona component in both bands, 
plus a very small additional contribution
to the fast core which is correlated but lagged by 0.2~days relative to HX 
from disk reverberation 
plus the broad correlated shoulder from propagation where SX leads. 
}

\rev{
These different components in SX are more clearly seen in the Fourier resolved lags in Fig. \ref{fig:sx_hx_ccf_frlag}b. 
The slow variability produced on the inner edge of this disc (soft X-ray excess component) propagates down and modulates 
the much faster variability of the hard X-ray corona, so the slow variability in HX is lagged by the propagation timescale.
By contrast, there is a small component of the fast variability in SX which is produced by reprocessing, so is lagged, 
but the lag timescale seen in the Fourier plot of $\sim 0.03$~days is very much shorter than the actual lag time of $\sim 0.2$~days
as it is diluted by the direct coronal component in the SX bandpass (see \citealt{Uttley14, Mizumoto18} for a discussion of dilution). }

\rev{There are several studies of the HX-SX Fourier lags (e.g. \citep{DeMarco13}). These use single, long, XMM-Newton, observations spanning $\sim 10^5$~s i.e. these start only at frequencies of $\sim 10^{-5}$~Hz, where our plots end. However, some of their AGN have much lower masses, so can sample this switch from propagation to reverberation. 
We note here that our model for the 
reverberating component is slightly different in that we assume that we are seeing mainly thermal reprocessed emission from X-ray heating of the inner edge of the disc, whereas these papers assume that the reprocessing is dominated by ionised relativistic reflection. Both components should be present at some level, but the expected signature from  reverberation lag is the same. 
We also note that our reverberation lag of 0.03~days is very much shorter than the true lag to the inner disc ($\sim 0.2$~days) due to dilution, and matches the very small lags seen in the data \citep{DeMarco13,kara16}.
}

\rev{
Overall our model gives a clear predictive behaviour of the X-ray variability, with a slow varying component propagating in from the disc 
giving a soft lead, while the fast variability intrinsic to the corona itself should imprint a (weak) reverberation signal producing a soft lag. 
This behaviour can now be  tested through X-ray monitoring campaigns of variable AGN, and in fact the soft X-rays leading the hard is already being seen in the case of Fairall 9 by Partington et al. (in prep.).
}


\bsp	
\label{lastpage}
\end{document}